\documentclass[aps,prl,reprint]{revtex4-2}

\usepackage{amssymb}
\usepackage{amsmath}
\usepackage{amssymb}
\usepackage{bm}
\usepackage{xcolor}
\usepackage{subcaption}
\usepackage{placeins}
\usepackage{float}
\usepackage[english=british]{csquotes}

\usepackage{graphicx} 
\usepackage{comment}
\usepackage{hyperref}

\renewcommand{\Pr}    {\mathrm{Pr}}
\newcommand{\Ra}    {\mathrm{Ra}}
\newcommand{\Nu}    {\mathrm{Nu}}
\newcommand{\Nuloc}    {\Nu_{\textrm{local}}}

\newcommand{\Smax}    {S_{\textrm{max}}}

\begin{document}
\setcounter{secnumdepth}{2}

\title{Lagrangian studies of coherent sets and heat transport in constant heat flux-driven turbulent Rayleigh-B\'enard convection}
\author{Philipp P. Vieweg}
\affiliation{Institute of Thermodynamics and Fluid Mechanics, Technische Universit\"at Ilmenau, 98693 Ilmenau, Germany}
\author{Anna Klünker}
\affiliation{Applied Mathematics, Leuphana Universit\"at L\"uneburg, 21335 L\"uneburg, Germany}
\author{J\"org Schumacher}
\affiliation{Institute of Thermodynamics and Fluid Mechanics, Technische Universit\"at Ilmenau, Germany}
\affiliation{Tandon School of Engineering, New York, 11201, NY, USA}
\author{Kathrin Padberg-Gehle}
\affiliation{Applied Mathematics, Leuphana Universit\"at L\"uneburg, 21335 L\"uneburg, Germany }
\date{\today}

\begin{abstract}
 We explore the mechanisms of heat transfer in a turbulent constant heat flux-driven Rayleigh-B\'enard convection flow, which exhibits a hierarchy of flow structures from granules to supergranules. Our computational framework makes use of time-dependent flow networks. These are based on trajectories of Lagrangian tracer particles that are advected in the flow. We identify coherent sets in the Lagrangian frame of reference as those sets of trajectories that stay closely together for an extended time span under the action of the turbulent flow. Depending on the choice of the measure of coherence, sets with different characteristics are detected. First, the application of a recently proposed evolutionary spectral clustering scheme allows us to extract granular coherent features that are shown to contribute significantly less to the global heat transfer than their spatial complements. Moreover, splits and mergers of these (leaking) coherent sets leave spectral footprints. Secondly, trajectories which exhibit a small node degree in the corresponding network represent objectively highly coherent flow structures and can be related to supergranules as the other stage of the present flow hierarchy. We demonstrate that the supergranular flow structures play a key role in the vertical heat transport and that they exhibit a greater spatial extension than the granular structures obtained from spectral clustering.  
\end{abstract}

\maketitle

\section{Introduction}
\label{sec:intro}
Turbulent thermal convection, the essential mechanism by which heat is transported in many natural flows \cite{Ahlers2009,Chilla2012},  has numerous geophysical \cite{Stevens2005}, astrophysical \cite{Schumacher2020} and technological \cite{Kelley2018} applications. Rayleigh-Bénard convection (RBC) is a paradigm of such natural thermal convection systems and consists basically of a fluid layer that is placed between two solid horizontal plates which uniformly heat and cool it from below and above, respectively \cite{Chilla2012}. Crucially, the formation of large coherent flow structures can be observed -- similar to the natural convection flows -- even in this simple system once a horizontally extended domain is provided \cite{Stevens2018,Pandey2018,Green2020,Krug2020,Vieweg2021a,Vieweg2022}. 
In the \textit{Eulerian frame of reference}, these long-living large-scale flow structures are termed (in the standard case where the plates possess spatially constant temperatures) turbulent superstructures as their characteristic horizontal scale exceeds the height of the convection layer. They consist of convection rolls and cells that are concealed in instantaneous velocity fields by turbulent fluctuations. However, they show up prominently after time-averaging of the velocity or temperature fields \cite{Pandey2018} and form the backbone of turbulent heat transport \cite{Fonda2019}. The situation is -- except for the formation of an entire hierarchy of different long-living large-scale flow structures -- roughly similar if the plates possess spatially constant temperature gradients or heat fluxes (more details follow further below) \cite{Vieweg2021a,Vieweg2022}.
The description of convection in the \textit{Lagrangian frame of reference} is connected to the \textit{material transport} and thus opens the perspective of a more precise quantification of the complex spatio-temporal pathways that heat takes through the fluid layer, including a classification into spatial regions that contribute more or less to its transfer. 

\begin{figure*}[t]
    \centering
    \includegraphics[scale=1.0]{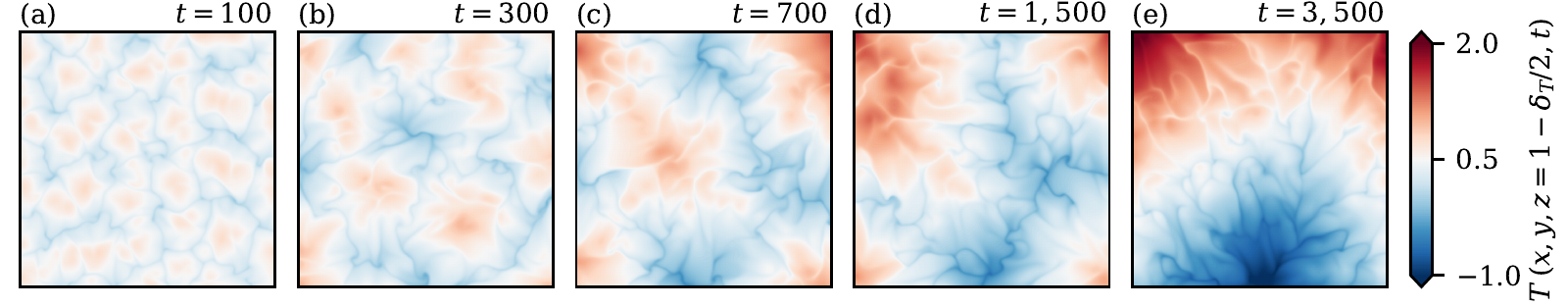}
    \caption{Rayleigh-Bénard convection, which is driven by an applied constant heat flux, exhibits a gradual supergranule aggregation. This time series visualizes instantaneous temperature fields $T$ close to the top plane (with $\delta_{T} = 1 / (2 \Nu)$ representing the thermal boundary layer thickness) for the simulation run that is at the focus of this present paper. The large and gradually growing patterns are termed \textit{supergranules}, whereas the smaller superposed structures are termed \textit{granules} and exhibit a time-\textit{in}dependent size. Note that the domain's horizontal extent is $30$ times its vertical extent and find more details on the configuration in section \ref{sec:problem}.}
    \label{fig:gradual_aggregation}
\end{figure*}

Lagrangian transport and mixing processes have been intensively studied by means of mathematical and computational tools from dynamical systems theory \cite{Allshouse2015,Haller2015,Hadjighasem2017} and the frameworks are built around different, but related notions and definitions of coherent flow structures. The concept of a finite-time coherent set \cite{FroylandLloydSan2010,Froyland2013,Karrasch2020}, a regularly shaped region in the fluid volume that only weakly mixes with its surrounding, is central to the present study. The boundaries of such regions can be identified within a geometric approach, where Lagrangian coherent structures represent minimal curves or surfaces that enclose coherent sets \cite{Haller2015}. 
Coherent sets were originally introduced within a probabilistic approach based on transfer operators \cite{FroylandLloydSan2010,Froyland2013} whereas recent methods make use of spatio-temporal clustering algorithms applied to Lagrangian trajectory data \cite{FroylandPadberg2015,Hadjighasem2016,Banisch2017,Schlueter2017,Padberg2017,Schneide2018,Wichmann2021,mowlavi_2022}. These aim at identifying coherent sets as groups of trajectories that remain close to each other during the time interval under investigation. Extensions have been proposed to study the long-term evolution -- including the emergence and decay of coherent sets \cite{Froyland2021,Schneide2022} --, while recent reviews discuss complex networks \cite{Iacobello2021,Donner2019} and machine learning \cite{Brunton2020} approaches for fluid flows.

Finite-time coherent sets in turbulent Rayleigh-Bénard convection, their relation to the Eulerian turbulent superstructures, and their role in heat transport have been studied using different Lagrangian approaches such as trajectory-based clustering \cite{Schneide2018,Schneide2019,Vieweg2021}, transfer operator methods \cite{Kluenker2020}, and evolutionary clustering \cite{Schneide2022}. In these works, aspect ratios (i.e., the ratio of the domain's horizontal to vertical extent) of $\Gamma \in \left[ 2, 16 \right]$ have been considered together with applied constant temperatures at the bottom and top plates (so-called Dirichlet boundary conditions). 
In contrast, when the convection is driven by a constant heat flux (representing Neumann boundary conditions), a gradual aggregation of smaller, but still large-scale convection cells -- termed \textit{granules} -- to even larger structures -- termed \textit{supergranules} -- is observed, see Figure \ref{fig:gradual_aggregation}. Note that the characteristic horizontal extension of the granular flow structures is $\mathcal{O} \left( H \right)$ (with $H$ being the layer height), whereas the supergranular structures grow until eventually being limited by the horizontal extent of the domain \cite{Vieweg2021a}. This growth can only be interrupted once the fluid layer is subject to additional physical mechanisms such as rotation around the vertical axis \cite{Vieweg2022}. However, this new mechanism has so far only been investigated in the Eulerian frame of reference \cite{Vieweg2021a,Vieweg2022}. The aim of the present paper is to extend the Lagrangian evolutionary framework \cite{Schneide2022} to this setting in order to address the following questions: 
\begin{enumerate}
\item How do Lagrangian coherent sets or features relate to granules and supergranules in the heat flux-driven convection case?
\item What is the role of granules and supergranules as the building blocks of the structural hierarchy on the global turbulent heat transport across the convection layer?
\item Can we observe the gradual supergranule aggregation in the Lagrangian frame of reference despite the superposition of different long-living large-scale flow structures?
\end{enumerate}
Particularly question 3 will require the evolutionary framework of the trajectory clustering analysis which we developed and outlined in \cite{Schneide2022} since the process will be affected by the slow transient aggregation.

The remainder of this paper is organized as follows. In Section \ref{sec:methods}, we briefly review the concept of Lagrangian coherent features in flows and the corresponding computational methods, focusing on the constructions introduced in \cite{Padberg2007,Schneide2022}. In Section \ref{sec:rbc}, we introduce the mathematical model for constant heat flux-driven Rayleigh-Bénard convection and briefly describe the results of our previous Lagrangian studies. Section \ref{sec:problem} provides objective Lagrangian characteristics of the flow and introduces subsequently methodical details of our following investigations. The results of the spectral clustering and the conditional heat transfer analysis are presented and discussed in Sections \ref{sec:results}  and \ref{sec:results1}, respectively. We conclude with a summary of the findings and an outlook in Section \ref{sec:conclusion}. 

\section{Lagrangian coherent features}
\label{sec:methods}

From the nonlinear dynamics point of view, there are a number of different concepts that describe the notion of Lagrangian coherent behavior in flows. Neglecting molecular diffusion, the motion of passive fluid particles is described by the ordinary differential equation 
\begin{equation}
\label{eq:advection_Lagrangian_particle}
\dot{\bm{X}} = \bm{u} \left( \bm{X}, t \right) .
\end{equation}
Here, $\bm{u} \left( \bm{X}, t \right)$ is a sufficiently smooth time-dependent velocity field with state $\bm{X} \in \mathbb{X}$, time $t \in \mathbb{R}$, and a compact subset $\mathbb{X} \subset \mathbb{R}^{\textrm{dim}}$ ($\textrm{dim} \in \left\{ 2, 3 \right\}$) which is the physical domain of the flow. Lagrangian particle trajectories are obtained as solutions $\bm{X} \left( t \right)$ of this differential equation \eqref{eq:advection_Lagrangian_particle}. Coherence can then be described in terms of the manner in which groups of Lagrangian trajectories behave. There are essentially three families of Lagrangian concepts for the identification of coherent features in a flow: geometric, probabilistic, and clustering approaches. Whereas the latter are data-based, the geometric and probabilistic methods have strong mathematical foundations in dynamical systems and ergodic theory. For discussions of the most frequently used frameworks, we refer to \cite{Allshouse2015,Hadjighasem2017}. We briefly review the different concepts below before we describe an evolutionary trajectory network (clustering) approach \cite{Schneide2022} that will be the basis for our studies. 

\subsection{Methods for identifying coherent behavior}
\textit{Geometric approaches} aim at identifying Lagrangian coherent structures, i.e. material surfaces that extremize a certain stretching or shearing quantity such as measured by the finite-time Lyapunov exponent. These material surfaces act as barriers to advective mass transport  \cite{Haller2015}. The classical concept has been recently extended to deal with weakly diffusive and stochastic transport \cite{Haller2018} and with transport of other quantities \cite{Balasuriya2018,Haller2020,Aksamit2023}. 
Lagrangian coherent structure methods require full knowledge of a smooth flow field, and so the extraction of evolving material surfaces in complex 3D flows remains challenging. 

In contrast, \textit{probabilistic methods} aim to identify full-volume finite-time coherent sets that minimally mix with the surrounding phase space. Coherent sets can be efficiently identified from $k$ leading singular vectors of sparse stochastic matrices that numerically approximate Perron-Frobenius (or transfer) operators, the latter of which describe the time evolution of densities in the phase or state space of the dynamical system. $k$ is determined based on a spectral gap criterion. The structures of interest are then extracted from the corresponding $k$ singular vectors via a hard clustering scheme such as $k$-means or a sparse eigenbasis approximation \cite{Froyland2019}. 

Related work considers Fokker-Planck equations \cite{Denner2016}, a dynamic Laplacian approach \cite{Froyland2015} with a finite-elements approximation \cite{FroylandJunge2018}, and a geometric heat flow construction \cite{Karrasch2020}. Bifurcations of coherent sets, such as their splits and mergers, have been studied both in transfer operator \cite{Blachut2020,Ndour2021} and dynamic Laplacian settings \cite{Froyland2021}. In \cite{Denes2022}, the different time scales of coherent sets have been studied.

While \textit{clustering} as a class of unsupervised machine learning algorithms allows to uncover underlying patterns in the data, different methods have already been proven to effectively extract coherent sets from Lagrangian trajectories. Given an appropriate similarity measure between trajectories (e.g. based on the average distance \cite{FroylandPadberg2015, Hadjighasem2016}), clustering results in groups of trajectories that are dynamically close to each other. First works in this context apply c-means clustering \cite{FroylandPadberg2015}, spectral clustering \cite{Hadjighasem2016,Banisch2017,Padberg2017}, graph coloring \cite{Schlueter2017}, density-based clustering \cite{Schneide2018,Wichmann2021,mowlavi_2022,Zeming2022} and complex networks \cite{Padberg2017,Banisch2019}. Interestingly, spectral clustering methods can be mathematically related to the transfer operator approaches \cite{Banisch2017}. Such methods allow us to extract and evaluate coherent sets for a given time window. To study, in contrast, the long-term evolution of coherent features -- including the emergence and decay of coherent sets --, it is necessary to relax their material properties \cite{Froyland2021,Schneide2022}. To this end, in \cite{Schneide2022} the trajectory-based network approach \cite{Padberg2017} has been extended to an evolutionary spectral clustering framework \cite{Chi2007} which blends a time-dependent analysis over shorter time spans with a dynamic short-term memory.

\subsection{Lagrangian trajectory networks and evolutionary spectral clustering}
\label{sec:networks}

In the following, we will briefly review the concept of trajectory networks and spectral clustering approaches, focusing -- due to their relevance to the present work -- on the constructions introduced in \cite{Padberg2017,Schneide2022}. 

Suppose we are given $N$ tracer trajectories $\bm{X}_i \left( t \right)$ with $i = 1 \ldots N$ at discrete time instances $t \in \mathbb{T} = \left\{ 0 \ldots \Theta \right\}$. Based on these trajectories, we construct an undirected network with Lagrangian particle trajectories serving as network nodes, where a link between two nodes is established if the respective trajectories come \enquote{close} to each other. There are different definitions of the distance between two trajectories and thus of \enquote{closeness}. In \cite{Padberg2017}, an unweighted network was considered with an adjacency matrix $A \in \left\{ 0, 1 \right\}^{N \times N}$, where $A_{ij}=1 $  if $\min_{t \in \mathbb{T}} \left\| \bm{X}_{i} \left( t \right) - \bm{X}_{j} \left( t \right) \right\| < \varepsilon$ for $i\neq j$ and  $A_{ij}= 0$ otherwise. $\varepsilon>0$ is thereby some given threshold. In \cite{Schneide2022,Weiland2023}, the number of $\varepsilon$-close encounters was further included as link weights by first constructing instantaneous adjacency matrices $A_{t} \in \left\{ 0, 1 \right\}^{N \times N}$ where $A_{ij, t} = 1$ if $\left\| \bm{X}_{i} \left( t \right) - \bm{X}_{j} \left( t \right) \right\| < \varepsilon$ for $i \neq j$ and $A_{ij, t} = 0$ otherwise, and then forming the symmetric network weight matrix $W = \sum_{t \in \mathbb{T}} A_{t}$, i.e., $W \in \mathbb{N}_0^{N \times N}$. In other words, an entry $W_{ij}$ is large if $\bm{X}_i$ and $\bm{X}_j$ are close in the sense that they have many $\varepsilon$-close encounters during the time span $\mathbb{T}$. 

\begin{figure}[t]
    \centering
    \includegraphics[width=0.7\columnwidth]{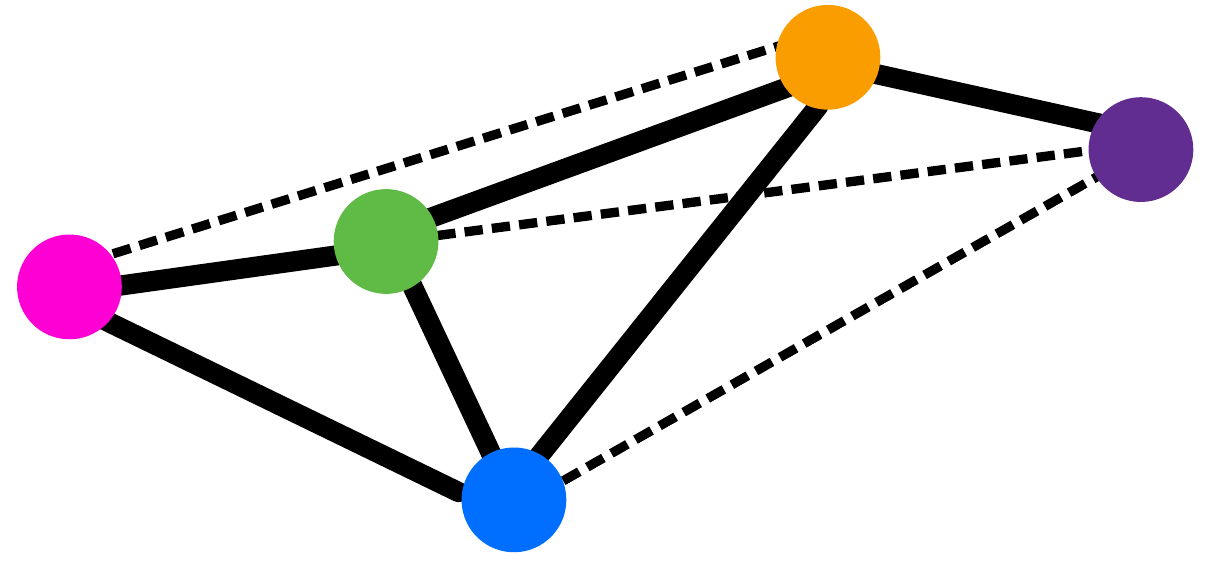}
    \caption{Illustration of the instantaneous network construction using mutual nearest neighbours. Here for each node its $K=3$ neighbours are linked (bold and dashed lines), but only those links are kept where the $K$-neighbourhood is mutual (bold lines). For instance, the $3$ nearest neighbours of the magenta node are blue, green and orange, however the magenta node does not belong to the $3$ nearest neighbours of the orange node. So the link between orange and magenta (dashed) is not considered.}
    \label{fig:schema_methode}
\end{figure}

As the choice of $\varepsilon$ is often tricky, especially when the given trajectory data is sparse, we propose and will use an alternative construction based on mutual $K$ nearest neighbours. Here, the instantaneous adjacency matrix has the entry $A_{ij, t} = 1$ if the particles $\bm{X}_i(t)$ and $\bm{X}_j(t)$ with $i \neq j$ are $K$ mutual nearest neighbours in the sense that $\bm{X}_i(t)$ belongs to the $K$ nearest neighbours of $\bm{X}_j(t)$ and simultaneously vice versa. Figure \ref{fig:schema_methode} provides an illustration of this idea. In this way, the link density as well as the connectivity can be controlled very conveniently by the choice of $K$. By including only mutual neighbourhoods, the adjacency matrices and thus the resulting weight matrix are symmetric just as in the $\varepsilon$-neighbourhood construction.  
We compare the different graph constructions based on $K$ and $\varepsilon$ in \ref{sec:graph_construction}. 

Node degrees are simple and useful network measures that evaluate the connections within the constructed graph. 
In our context, the unweighted node degree $d_{i}$ is defined by the number of different trajectories that are linked to node ${\bm X}_i$. That is, $d_{i}=\sum_{j=1}^N \overline{W}_{ij}$ where $\overline{W} \in \{0, 1\}^{N\times N}$ is the adjacency matrix of the network, encoding only the presence of links and thus being related to $W$. This network measure can be related to the finite-time Lyapunov exponent \cite{Banisch2019} as well as to the trajectory encounter volume \cite{Rypina2017}.

To identify coherent sets from the weighted trajectory network, we make use of \textit{spectral} clustering \cite{ShiMalik2000, Bach2006} with the weight matrix $W$ serving as a similarity matrix. In order to partition the trajectories into $k$ groups, we aim to subdivide the graph such that the within-cluster similarity is maximized whereas the between-cluster similarity is simultaneously minimized. The solution to this is a $k$-way normalized cut. These optimization problems as formulated in \cite{ShiMalik2000, Bach2006} are NP-hard, but spectral relaxations have been proposed. We follow \cite{Bach2006} and minimize the cost function
\begin{equation}
\label{eq:cost_function}
{\rm Cost}_{\textrm{NCut}} = k - {\rm Tr} \left[ \Xi^{\textrm{T}} \left( D^{-1/2} W D^{-1/2} \right) \Xi \right] 
\end{equation}
over all orthogonal matrices $\Xi \in \mathbb{R}^{N \times k}$. $D \in \mathbb{R}^{N \times N}$ represents a diagonal matrix with $D_{ii} = \sum_{j=1}^N W_{ij}$. This minimization problem is solved by the matrix $\Xi$ if it consists of the eigenvectors corresponding to the $k$ largest eigenvalues of the matrix $D^{-1/2} W D^{-1/2}$ as columns. Complementing the nodes' coordinates captured in these eigenvectors, gaps in the corresponding eigenvalue spectrum  provide information on the intrinsic connectivity of the graph \cite{Banisch2017} and thus the potential numbers of coherent sets $k$ \cite{Schneide2022} in a data-driven way. In our context, we use a simple spectral gap criterion, guided by the physical knowledge of the length scales of structures of interest. Different heuristics how to choose an appropriate $k$ without \emph{a priori} knowledge can be found e.g.\ in \cite{Froyland2019,Filippi2021}. We note that the choice of $k$ also effectively fixes the size of the structures to be identified.

To extract $k$ coherent sets from these coordinates in eigenspace, we post-process the eigenvectors using the \textit{sparse eigenbasis approximation} (SEBA) algorithm \cite{Froyland2019}. In a nutshell, this algorithm transforms the set of eigenvectors $\Xi$ to a new set of vectors $S \in \mathbb{R}^{N \times k}$ which spans approximately the same subspace but is significantly sparser. As different coherent features or sets are now separated into different sparse vectors (as the columns of $S$), the entry $S_{ij}$ can be interpreted as the likelihood of trajectory $\bm{X}_{i}$ to belong to cluster $j$. In this work, we are only interested in the maximum likelihood of cluster affiliation of each trajectory (to belong to any cluster) and so we define subsequently the cluster membership indicator $S_{\textrm{max, } i} = \max_{j} S_{ij}$.

In \cite{Schneide2022}, such a network-based approach was extended to study the evolution of large-scale coherent sets in turbulent flows over long time spans. We briefly review the evolving cluster approach as the one of the two proposed frameworks which will be used in the present study. For more details, we refer to \cite{Schneide2022}.

The evolving cluster approach is based on the preserving cluster membership framework from \cite{Chi2007}. Let $W_t$ denote the time-dependent weight matrices that are now computed over a shorter observation window of length $\Delta t_{\textrm{ow}} \ll \Theta$ and centered around some time $t \in \mathbb{T}$ -- more precisely, $W_t = \sum_{s \in \mathbb{T}_{t}, \Delta t_{\textrm{ow}}} A_{s}$ with $\mathbb{T}_{t, \Delta t_{\textrm{ow}}} = \left\{ t - \Delta t_{\textrm{ow}} / 2, \ldots ,  t + \Delta t_{\textrm{ow}} / 2 \right\}$ for times $t \in \mathbb{T}$. Now, for each $t$, we minimize (cf. \cite{Chi2007})
\begin{equation}
\label{eq:cost_function_t0}
{\rm Cost}_{\textrm{NCut, } t} = k - {\rm Tr} \left[ \Xi_t^{\textrm{T}} \widehat{W}_t \Xi_{t} \right]
\end{equation}
which is (again) solved by $\Xi_{t} \in \mathbb{R}^{N \times k}$ if it contains the eigenvectors corresponding to the $k$ largest eigenvalues of the matrix 
\begin{equation}
\label{eq:evocl}
\widehat{W}_{t} = \mu D_t^{-1/2} W_t D_t^{-1/2} + \left( 1 - \mu \right) \Xi_{t - \Delta t_{\textrm{s}}} \Xi_{t - \Delta t_{\textrm{s}}}^{\textrm{T}} .
\end{equation}
Note here that $\Xi_{t - \Delta t_{\textrm{s}}}$ represent the eigenvectors obtained from the clustering at the previous time instance $t-\Delta t_{\textrm{s}}$, where $\Delta t_{\textrm{s}}$ denotes the time shift. The obliviousness parameter $\mu$ regulates the importance of the current connectivity of the graph compared to the previous clustering, and using $\mu = 1$ would lead to the usual clustering over the current time interval $\left[ t - \Delta t_{\textrm{ow}} / 2, t + \Delta t_{\textrm{ow}} / 2 \right]$, similar to eq. \eqref{eq:cost_function}. 
However, note that $\widehat{W}_{t}$ is (once $\mu < 1$) no longer sparse due to $\Xi_{t - \Delta t_{\textrm{s}}} \Xi_{t - \Delta t_{\textrm{s}}}^{\textrm{T}}$ being a dense matrix. To overcome this issue, we use instead the sparse approximation $S_{t - \Delta t_{\textrm{s}}}$ of the eigenspace spanned by $\Xi_{t - \Delta t_{\textrm{s}}}$ (see also \cite{Schneide2022} for more details). 

In the following, we denote by $\Smax \left[ \bm{X}_i \left( t \right) \right]$ the cluster membership indicator for the trajectory fragment $\bm{X}_i \left( t \right)$ over the time interval $\mathbb{T}_{t, \Delta t_{\textrm{ow}}}$ (i.e.\ centred at $t$), similarly for the node degree $d \left[ \bm{X}_i \left( t \right) \right]$. Note in particular that -- despite the inclusion of historical information -- $d \left[ \bm{X}_i \left( t \right) \right]$ is still obtained from $W_t$ (instead of $\widehat{W}_t$).

\section{Turbulent Rayleigh-Bénard convection}
\label{sec:rbc}
In this section, we introduce the present configuration of Rayleigh-Bénard convection and briefly review the results of previous Lagrangian studies.

\subsection{Model definition and computational details}
We study Lagrangian coherent sets in a three-dimensional Cartesian domain using the model setting introduced in \cite{Vieweg2021a}. Therefore, the equations of motion are made dimensionless based on characteristic quantities of the system. In addition to the layer height $H$ as unit of length, we adopt the free-fall time scale $\tau_{\textrm{f}} = H / U_{\textrm{f}}$ as the unit of time together with the corresponding free-fall velocity $U_{\textrm{f}} = \sqrt{\alpha g T_{\textrm{char}} H}$. Here, $\alpha$ represents the isobaric expansion coefficient and $g$ the acceleration due to gravity. $T_{\textrm{char}}$ is a characteristic temperature that depends on the thermal boundary conditions and thus on either the prescribed vertical temperature gradient $\beta > 0$ or the applied temperature difference $\Delta T > 0$ between the plates. Hence, the governing equations follow in their non-dimensional description as
\begin{align}
\label{eq:CE}
\nabla \cdot \bm{u} &= 0 , \\
\label{eq:NSE}
\frac{\partial \bm{u}}{\partial t} + (\bm{u} \cdot \nabla ) \thickspace \bm{u} &= -\nabla p + \sqrt{\frac{\Pr}{\Ra}} \nabla^{2} \bm{u} +T \bm{e}_{z} , \\
\label{eq:EE}
\frac{\partial T}{\partial t} + ( \bm{u} \cdot \nabla ) \thickspace T &= \frac{1}{\sqrt{\Ra\Pr}} \nabla^{2} T,
\end{align}
where $p$, $\bm{u}$, and $T$ are the pressure, velocity, and temperature field, respectively. Two dimensionless parameters -- the Prandtl number $\Pr$ and the Rayleigh number $\Ra$ -- control the entire dynamics and summarize all the dimensional coefficients and parameters. These numbers are defined as
\begin{equation}
\label{eq:def_Pr_RaD_RaN}
\Pr = \frac{\nu}{\kappa}  \thickspace \thickspace \thickspace
\textrm{ and} \thickspace \thickspace \thickspace
\Ra= \frac{\alpha g T_{\textrm{char}} H^{3}}{\nu \kappa} 
\end{equation}
with the kinematic viscosity and temperature diffusivity of the fluid, $\nu$ and $\kappa$, respectively.

These equations are complemented by the numerical domain and its corresponding boundary conditions. With the aspect ratio as the ratio of the domain's horizontal to vertical extent, $\Gamma = L / H$, the domain extends (non-dimensionally in units of the layer height $H$) across $x,y \in \left[ - \Gamma / 2, \Gamma / 2 \right]$ and $z \in \left[ 0, 1 \right]$. The impermeable bottom and top planes at $z \in \left\{ 0, 1 \right\}$ obey free-slip mechanical boundary conditions 
\begin{equation}
u_{z} = 0 \quad\mbox{and}\quad \frac{\partial u_{x}}{\partial z} = \frac{\partial u_{y}}{\partial z} = 0 ,
\end{equation}
whereas the lateral boundaries are closed and thermally insulating (walls), i.e., 
\begin{equation}
u_{n} = 0, \quad \frac{\partial u_{t}}{\partial n} = 0, \quad \textrm{and} \quad \frac{\partial T}{\partial n} = 0 
\end{equation}
with $n$ and $t$ being the corresponding wall-normal and -tangential coordinates, respectively. In contrast to the classical scenario (as in \cite{Vieweg2021}) where one applies constant temperatures at the top and bottom planes such that $T \left( z = 0 \right) = 1$ and $T \left( z = 1 \right) = 0$ with $T_{\textrm{char}} = \Delta T$, we prescribe here a \textit{constant vertical temperature gradient} 
\begin{equation}
\left. \frac{\partial T}{\partial z} \right|_{z = 0} = \left. \frac{\partial T}{\partial z} \right|_{z = 1} = - 1 .
\end{equation}
The characteristic temperature is thus $T_{\textrm{char}} = \beta H$. Note that in this present case, the temperature at the top and bottom planes is allowed to vary locally -- all that is fixed is the applied constant heat flux through the spatially constant temperature gradient.

We solve the equations of motion using the spectral element solver \textsc{nek5000} \citep{Fischer1997,Scheel2013} and resolve all dynamically relevant scales down to the Kolmogorov scale $\eta_{\textrm{K}}$. Any simulation starts with the fluid at rest and the linear conduction temperature profile, the latter of which is randomly and infinitesimally perturbed. The temporal advancement of the simulations exploits a second order backwards difference formula. More detailed information on numerical details can be found in our previous works \cite{Scheel2013, Pandey2018, Vieweg2021a, Vieweg2021,Fonda2019, Vieweg2022}.

Finally, the Nusselt number $\Nu$ is the \textit{global} measure to quantify the relevance of convective heat transfer across the fluid layer. For constant heat flux boundary conditions, it is given by \cite{Otero2002}
\begin{equation}
\label{eq:def_NuN}
\Nu = \frac{1}{\langle T \left( z = 0 \right) \rangle_{A} - \langle T \left( z = 1 \right) \rangle_{A}} 
\end{equation}
with $A = \Gamma \times \Gamma$ being the non-dimensional horizontal cross section.

While the spectral element code works in the Eulerian frame of reference, the advection of $N$ Lagrangian and thus massless tracer particles is an analysis in the Lagrangian frame of reference. These tracers follow the surrounding flow perfectly and so their trajectories $\bm{X}_{i} \left( t \right)$ are given by
\begin{equation}
\label{eq:Lag}
\dot{\bm{X}_{i}} = \bm{u} \left( \bm{X}_{i}, t \right) \quad \textrm{for } \thickspace i = 1 \dots N ,
\end{equation}
see again eq.\ \eqref{eq:advection_Lagrangian_particle}.
The velocity field components in this equation are therefore obtained from a spectral interpolation of the Eulerian fields to the tracer position $\bm{X}_{i} \left( t \right)$. Hence, the spatial aspect of the advection of the Lagrangian particles is as accurate as the simultaneous temporal advancement of the Boussinesq equations \eqref{eq:CE} -- \eqref{eq:EE} in the Eulerian frame of reference. Concerning the temporal aspect of this particle advection, a third-order Adams-Bashforth time stepping is exploited.

In addition to the global measure of turbulent heat transport (see above), we compute for each individual tracer particle a \textit{local} Nusselt number \cite{Vieweg2021}
\begin{equation}\label{eq:localNu}
\widehat{\Nu}_{\textrm{local}} \left[ \bm{X}_{i} \left( t \right) \right] = \left. \sqrt{{\rm Ra} {\rm Pr}} \thickspace u_{z} T \vphantom{\frac{\partial T}{\partial z}} \right|_{\bm{X}_{i} \left( t \right)} - \left. \frac{\partial T}{\partial z} \right|_{\bm{X}_{i} \left( t \right)} \
\end{equation}
in the Lagrangian frame of reference along its trajectory $\bm{X}_{i} \left( t \right)$. To quantify eventually the overall heat transport of a trajectory fragment over the time set $\mathbb{T}_{t, \Delta t_{\textrm{ow}}}$ centered around $t$, we compute the time-averaged local Nusselt number
\begin{equation}
\label{eq:localNu_av}
\Nuloc \left[ \bm{X}_{i} \left( t \right) \right] = \langle \widehat{\Nu}_{\textrm{local}} \left[ \bm{X}_{i} \left( t \right) \right] \rangle_{\mathbb{T}_{t, \Delta t_{\textrm{ow}}}}.
\end{equation}
Note that the definition of the local Nusselt number is the same as in \cite{Vieweg2021} even for the different thermal boundary conditions once the fields are re-scaled based on $\Nu$.

\subsection{Previous Lagrangian studies}
Unlike in the present work where convection is driven by a constant heat flux, our previous Lagrangian analyses have been restricted to the setting of prescribed constant temperatures at the top and bottom boundaries. We summarize our major findings from these past studies in the following.

A first numerical analysis of turbulent superstructures in a large-aspect-ratio $3$-dimensional convection cell by means of Lagrangian particle trajectories was carried out in \cite{Schneide2018} with $\Pr=0.7$, $\Ra=10^5$, and $\Gamma = 16$. Coherent sets have been detected from spectral clustering of a weighted and undirected trajectory network from the trajectory points of Lagrangian particles while link weights have been determined by a mean dynamical distance between different particle trajectories. It was demonstrated that the resulting Lagrangian trajectory clusters, which were obtained by a subsequent $k$-means clustering, are related to bigger patches of the turbulent superstructures from the Eulerian frame of reference. Furthermore, the characteristic Lagrangian time and length scales of the superstructures were found to agree very well with their Eulerian counterparts. For longer time periods beyond the characteristic time, a density-based clustering (DBSCAN \cite{Ester1996}) by means of time-averaged Lagrangian pseudo-trajectories was applied. A small coherent subset of the pseudo-trajectories has been obtained consisting of those Lagrangian particles that are trapped for long times in the core of the circulation rolls and are thus not subject to ongoing turbulent dispersion.

In \cite{Schneide2019}, different trajectory-based Lagrangian methods to identify coherence in a simple $2$-dimensional Rayleigh-Bénard benchmark problem with $\Pr = 10$, $\Ra=10^6$, and $\Gamma = 4$ have been compared and found to give consistent results. By introducing the concept of heat particles, we extended the notion of coherent sets to include heat transport. Our findings showed that convective rolls, which were identified by Lagrangian coherent sets, contribute least to the vertical heat transport. Furthermore, \cite{Kluenker2020} demonstrated that also the applications of transfer operator and dynamic Laplacian approaches give consistent results in $2$- and $3$-dimensional RBC model systems and compare well with the structures identified in previous investigations \cite{Schneide2018,Schneide2019}. 

Lagrangian heat transport in a complex large-aspect ratio $3$-dimensional RBC system was first analysed and quantified in \cite{Vieweg2021}. Here we considered a set of three different fluids with $\Pr \in \left\{ 0.1, 0.7, 7 \right\}$ at $\Ra=10^5$ and $\Gamma = 16$ (one of these runs as in \cite{Schneide2018}). A diffusion distance, similar in construction to \cite{Banisch2017}, was used to set up the network weight matrices and allowed us after applying a sparse eigenbasis approximation eventually to conclude that Lagrangian coherent sets contribute (independently of the fluid) significantly less -- more precisely, up to one third less -- to the vertical heat transport than their spatial counterparts. 

While our approach in \cite{Vieweg2021} accounted for the slow re-organisation of the turbulent superstructures over longer times only by repeating the analysis of relatively short time windows for several distinct times,\cite{Schneide2022} was eventually able to address the long-term dynamics of coherent sets in both $2$- and $3$-dimensional RBC simulations. In particular, we analysed a simple $2$-dimensional setup at $\Pr=7$, $\Ra=10^8$, and $\Gamma=8$ as well as a complex $3$-dimensional setup at $\Pr=0.7$, $\Ra=10^5$, and $\Gamma=16$ (the latter as in \cite{Schneide2018}). The proposed evolutionary clustering framework eases the mathematical rigor of material transport and allows to monitor the long-term dynamics of Lagrangian coherent sets including their splits or mergers. These events were traced back to the changes of spectral properties of the corresponding network matrices. This study confirmed once more that coherent sets effectively suppress the turbulent heat transport across the domain, with their contribution to the turbulent heat transfer being reduced by about 30\% in the two-dimensional and 13\% in the three-dimensional case, respectively. This evolutionary trajectory clustering framework from \cite{Schneide2022} represents -- together with the physical insights obtained from \cite{Vieweg2021a} -- the basis for our current study.

\section{Lagrangian characteristics of the present convection flow and the initial graph construction}
\label{sec:problem}

\begin{figure}[t]
    \centering   
    \includegraphics[scale = 1.0]{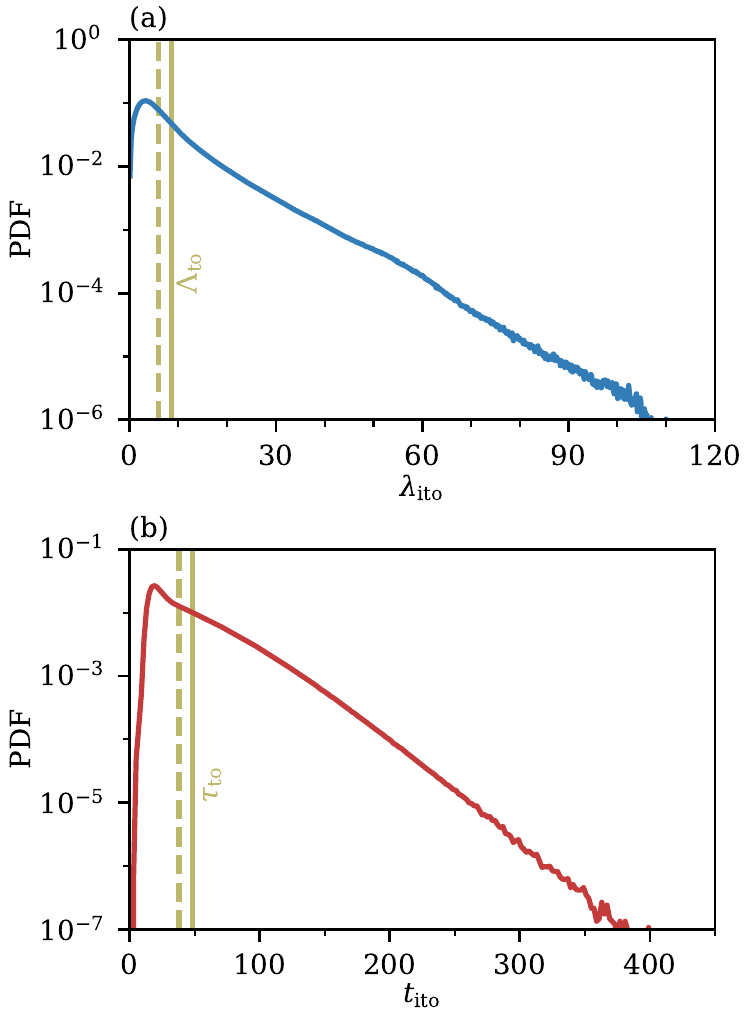}
    \caption{Distribution of the individual turnover length and time scales, $\lambda_{\textrm{ito}}$ and $\tau_{\textrm{ito}}$, respectively, of the Lagrangian tracer particles in the flow. The vertical dashed and solid lines in both panels represent the median and the mean values, respectively. These provide estimates for the characteristic Lagrangian turnover length and time scales, $\Lambda_{\rm to}$ and $\tau_{\rm to}$ (measured in layer heights $H$ and free-fall times $\tau_{\rm f}$), respectively.} 
    \label{fig:character}
\end{figure}

At the centre of this present study is simulation run Nfs1 from \cite{Vieweg2021a} at $\Pr = 1$ and $\Ra = 10,430$, which we repeat here in a laterally closed $\Gamma = 30$ cell (being in contrast to the periodic $\Gamma = 60$ domain in \cite{Vieweg2021a}). As introduced in sections \ref{sec:intro} and \ref{sec:rbc}, this flow is driven by a constant heat flux and, consequently, exhibits a hierarchy of different long-living large-scale flow structures. In more detail, the \textit{granules} -- as the first stage in this hierarchy -- offer a time-independent size of $\Lambda_{\textrm{G}} \sim \mathcal{O} \left( 1 \right)$, whereas the \textit{supergranules} -- as the second stage -- grow gradually over time with a final statistically stationary size of $\Lambda_{\textrm{SG}} \gg 1$. In a horizontally periodic domain, $\Lambda_{\textrm{SG}} = \Gamma$ in the end. For the present setup, we can quantify the time-independent characteristic wavelength of the granular flow structures $\Lambda_{\textrm{G}} \approx 5 \dots 6$ via the corresponding spectral peak in the vertical velocity field close to the top plane. Although the simulation covers in total $3.5 \times 10^{3} \tau_{\textrm{f}}$ with a simultaneous advection of $512^{2} \times 2$ Lagrangian particles, we restrict our subsequent analysis to $256^{2}$ Lagrangian trajectories during the first $1 \times 10^{3} \tau_{\textrm{f}}$ at a temporal resolution of $1 \tau_{\textrm{f}}$. This will capture the major part of the aggregation process, see again Figure \ref{fig:gradual_aggregation}.

\begin{figure*}[t]
    \centering
    \includegraphics[scale = 1.0]{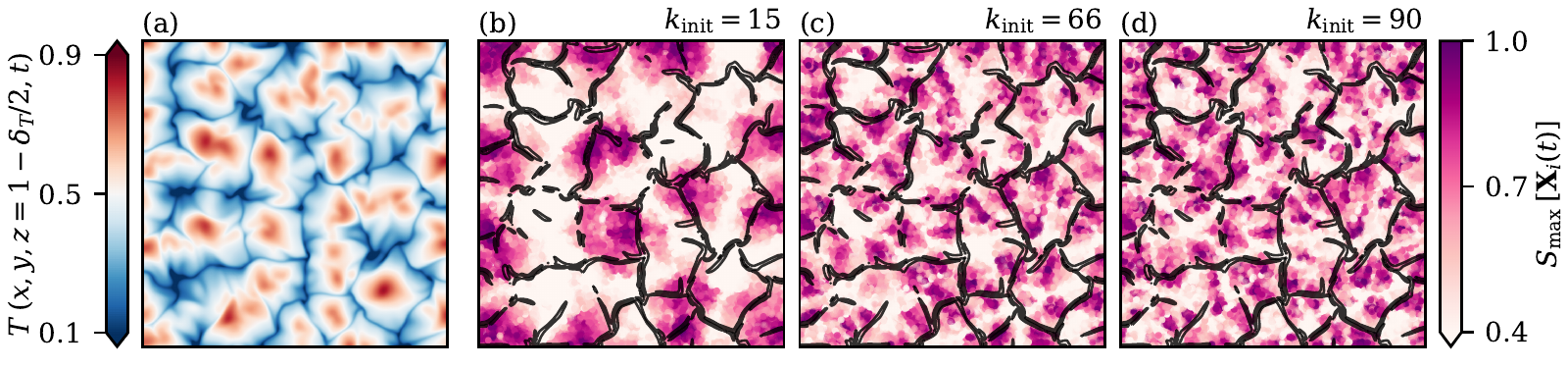}
    \caption{The initial observation window at $t = 124$ exhibits (a) the ongoing supergranule aggregation in its temperature field $T$ and (b -- d) allows to extract different coherent features as shown via the maximum likelihood of cluster affiliation $\Smax$ as an output of the SEBA post-processing for different choices of initially present features $k_{\textrm{init}}$. Note that these numbers are suggested by gaps in the initial eigenspectrum as shown in Figure \ref{fig:spectrum}. The black contour lines in panels (b -- d) correspond to the iso-level $\chi=((\partial T / \partial x)^{2} + ( \partial T / \partial y)^{2})^{1/2} = 0.4$ and encode regions of strong down-welling fluid, the latter of which can be expected to separate coherent flow regions.}
    \label{fig:initial_clusters_at_different_kinit}
\end{figure*}

Lagrangian characteristics provide objective physical or data-driven insights into the actual flow at hand, and help eventually to set many of the hyper-parameters related to the clustering procedure. For this reason, we determine the characteristic turnover length and time scales of the flow, $\Lambda_{\textrm{to}}$ and $\tau_{\textrm{to}}$, respectively. To obtain these characteristics, we use the entire dataset of $512^{2} \times 2$ particles across $3.5 \times 10^{3} \tau_{\textrm{f}}$.
On the one hand, the characteristic turnover wavelength $\Lambda_{\textrm{to}}$ is determined from the distribution of the individual turnover lengths $\lambda_{\textrm{ito}}$, the latter of which represent four times the horizontal travel distance of every particle between two successive intersections of the midplane $z = 0.5$. Depending on the chosen typical measure (mean or median) of the resulting probability density function (PDF), see Figure \ref{fig:character} (a), this results in $\Lambda_{\textrm{to}} = 5.8 \dots 8.5$ and agrees thus very well with $\Lambda_{\textrm{G}}$.
On the other hand, the characteristic turnover time $\tau_{\textrm{to}}$ is similarly derived from the statistics of the individual turnover times $t_{\textrm{ito}}$, the latter of which are probed by the particles passing the horizontal planes $z = 0.2$ and $z = 0.8$ to complete an entire turnover movement. Again, depending in the chosen typical measure, this yields $\tau_{\textrm{to}} = 38 \dots 48$, see Figure \ref{fig:character} (b).

We start our Lagrangian analysis at $100 \tau_{\textrm{f}} \approx 2 \dots 3 \tau_{\textrm{to}}$ to allow for a proper distribution of the particles across the entire domain after the initial seeding. We set up our instantaneous adjacency matrices $A_t$ (see Section \ref{sec:networks}) by using $K = 26$ mutual nearest neighbours. \ref{sec:graph_construction} discusses and compares different choices of parameters for the network construction. 

Based on $\tau_{\textrm{to}}$, we use here observation windows of width $\Delta t_{\textrm{ow}} = 48 \approx \tau_{\textrm{to}}$ and so the first window is centered around $t = 124$ -- Figure \ref{fig:initial_clusters_at_different_kinit} (a) visualizes the corresponding instantaneous temperature field close to the top plane.

\begin{figure}[t!]
    \centering
    \includegraphics[scale = 1.0]{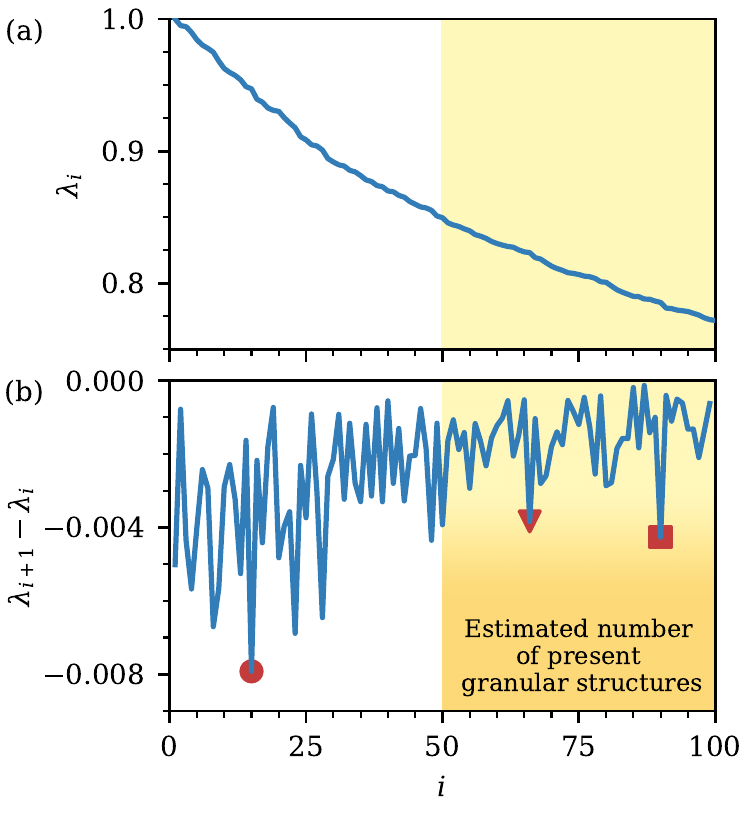}
    \caption{During the first observation window at $t = 124$, (a) the spectrum of the leading $100$ eigenvalues $\lambda_{i}$ of the graph displays several gaps, the latter of which are highlighted in panel (b). In accordance with the  estimated number of granular flow structures, which is based on the characteristic turnover wavelength $\Lambda_{\textrm{to}}$, we pick the gap at $k_{\textrm{init}} = 66$ (filled triangle) for the analysis in the main text. The evolution of structures corresponding to two further prominent gaps at $k_{\textrm{init}} = 15$ (filled circle) and $k_{\textrm{init}} = 90$ (filled square) are analysed and contrasted in \ref{sec:fifteensets}.}
    \label{fig:spectrum}
\end{figure}

As there is no historical information available yet, $\mu = 1$ for this very first observation window and so we compute the leading $100$ eigenpairs of only $D_t^{-1/2} W_t D_{t}^{-1/2}$ in eq. \eqref{eq:evocl}. The corresponding eigenspectrum, see Figure \ref{fig:spectrum} (a), offers beside the most prominent spectral gap at $k_{\textrm{init}} = 15$ a vast number of further potential gaps as highlighted by Figure \ref{fig:spectrum} (b) -- this circumstance makes an estimation of the present number of structures of interest vital to any further analysis. Presuming the Lagrangian turnovers from above to be predominantly caused by the granular flow structures as the first stage in the hierarchy -- which (i) should complete a turnover more quickly than even larger flow structures and (ii) is supported by $\Lambda_{\textrm{to}} \approx \Lambda_{\textrm{G}}$  --, the extracted characteristics allow to estimate the number of present granules to $\left[ \Gamma / \left( \Lambda_{\textrm{to}} /2 \right) \right]^{2} \approx 50 \dots 100$. One can detect two further prominent spectral gaps within this range as highlighted by additional markers in Figure \ref{fig:spectrum} (b). 

In order to decide for one of these potential numbers of coherent features, $k_{\textrm{init}} = \left\{ 15, 66, 90 \right\}$, we visualize all of them in Figure \ref{fig:initial_clusters_at_different_kinit} (b -- d). Our past studies \cite{Vieweg2021, Schneide2022} showed that up- and down-welling fluid separates coherent spatial regions, and so one should expect that the coherent features here are again influenced or separated by such regions -- we thus superimpose the regions of strongly down-welling fluid as black contour lines. Together with our estimation of the number of present granular flow structures from above, this discards $k_{\textrm{init}} = 15$ as a proper choice. In contrast, both $k_{\textrm{init}} = \left\{ 66, 90 \right\}$ seem to extract features that are influenced by the incoherent down-welling fluid.
Thus we select the first of these peaks, $k_{\textrm{init}} = 66$, as the initial number of coherent sets for the upcoming evolutionary clustering. We will contrast the results for all of the $k_{\textrm{init}} = \left\{ 15, 66, 90 \right\}$ in \ref{sec:fifteensets}.

Together with these mostly data-driven or physics-informed decisions of several hyper-parameters, we move in the next section on to incorporate historical information in the evolutionary spectral clustering approach. First, we will evaluate the coherence of long-living large-scale flow structures based on the maximum likelihood of cluster affiliation $\Smax$ as an output of the evolutionary SEBA post-processing -- this quantity is influenced by the incorporation of historical information. Second, we will contrast these results with the current structure of the network weight matrix in terms of the node degree $d$. Recall here that while this quantity does not take any historical information into account, it simultaneously does not require to solve the expensive eigenvalue problem. Finally, note in particular that high coherence implies $\Smax \left[ \bm{X}_{i} \left( t \right) \right] \geq S_{\textrm{max, ref}}$ or $d \left[ \bm{X}_{i} \left( t \right) \right] \leq d_{\textrm{ref}}$, where the respective reference values are deduced from ensemble means as outlined below.

\section{Evolutionary spectral clustering analysis}
\label{sec:results}

In order to incorporate historical information from  the previously extracted coherent sets in each subsequent clustering instance, one needs to shift the observation window by some small amount $\Delta t_{\textrm{s}}$ and simultaneously incorporate historical information based on the obliviousness parameter $\mu$ -- see again eq. \eqref{eq:evocl}. Based on a preliminary variation of $\mu \in \left\{ 1.0, 0.95, 0.8 \right\}$ given a fixed $\Delta t_{\textrm{s}} = 2$, see \ref{sec:different_mu}, we proceed here in the main text with $\mu = 0.95$.
While we continue to use a spectral gap criterion to detect the actually present number of coherent features, we allow this number only to vary by $\pm 1$ with each subsequent clustering instance. This allows to automate the entire remaining evolutionary approach. Given these choices of parameters and the features extracted from the initial graph at $t=124$, we thus automatically construct $\widehat{W}_t$ for $t = \left\{ 126, 128, \dots \right\}$ and obtain in this way $427$ consecutive observation windows.

\begin{figure*}[t!]
    \centering
    \includegraphics[scale = 1.0]{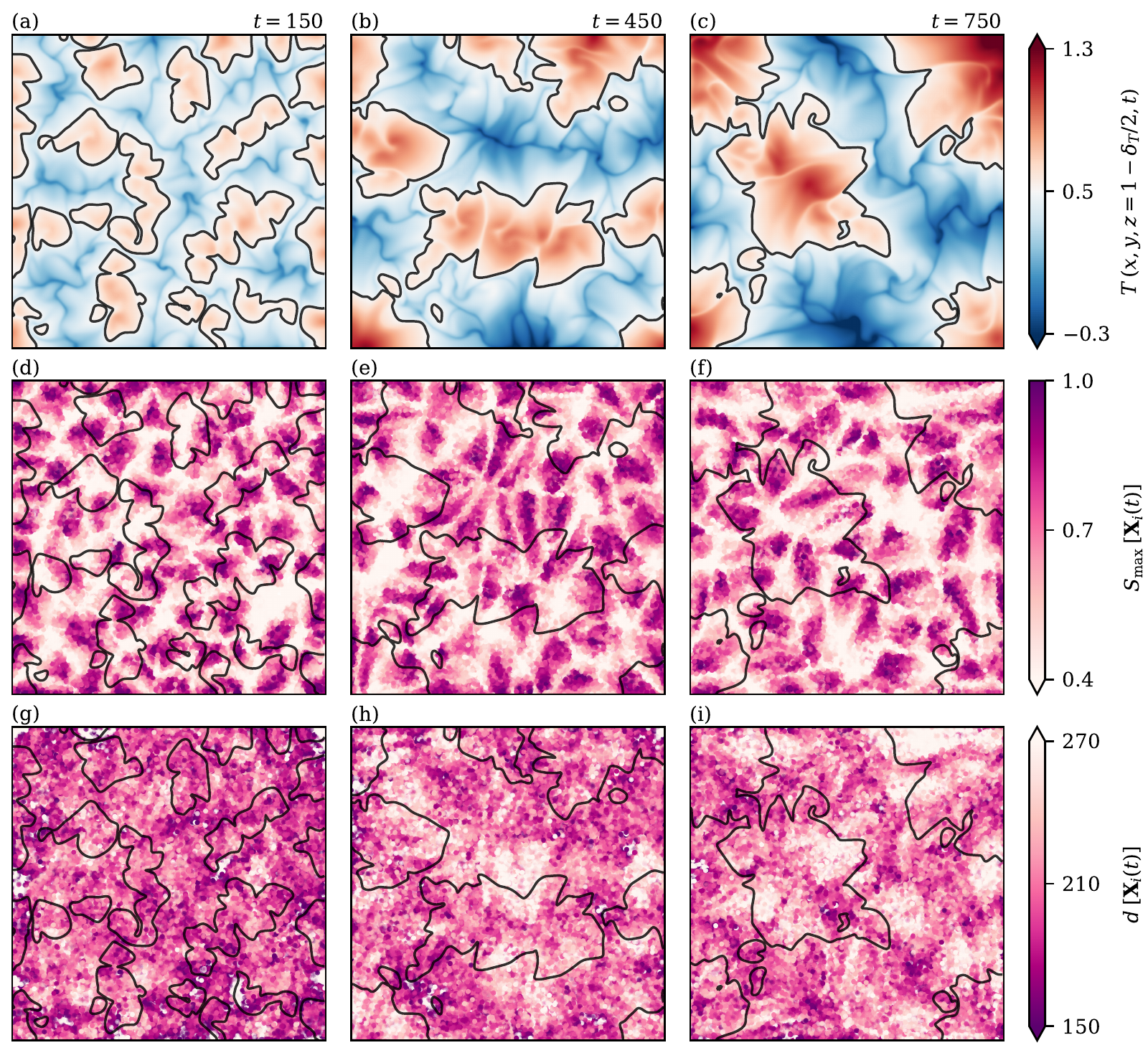}
    \caption{Evolution of coherence during the gradual supergranule aggregation. 
    (a -- c) The temperature field indicates the transient process.
    (d -- f) The maximum likelihood of cluster affiliation $\Smax$ tracks the evolution of the different evolving granular coherent features.
    (g -- i) In contrast, the (unweighted) node degree seems to capture even the gradual supergranule aggregation as the other stage in the flow hierarchy.
    Note that the colormaps in panels (d -- i) are chosen such that more intense colors express stronger coherence. The superimposed black lines contrast the different measures of coherence with the $T = 0.5$ temperature contours based on panels (a -- c) and indicate the ongoing supergranule aggregation.}
    \label{fig:smax66}
\end{figure*}
While the initial number of coherent features is motivated by the number of present \textit{granular} flow structures as the first stage of the present hierarchy of granular and supergranular flow structures, these features are subsequently influenced by the previous clustering and allowed to evolve over time. Consequently, there may appear mergers, splits, or even establishments of completely new ones. 
In order to categorize the extracted evolving coherent sets or features with respect to the present hierarchy, we start by visualizing a time series of the instantaneous temperature field (which prominently indicates the gradual supergranule aggregation \cite{Vieweg2021a,Vieweg2022}) in Figure \ref{fig:smax66} (a -- c) together with the maximum likelihood of cluster affiliation $\Smax$ in panels (d -- f). Note in case of the latter that regions of more intense color correspond to aspects of the flow with stronger coherence, whereas less intensely colored regions indicate some incoherent background. We superimpose the $T = 0.5$ contours of the temperature fields from panels (a -- c) as solid lines to contrast the gradual growth of the \textit{supergranules} with the evolving features. Although the locations of the coherent sets seem -- to some extent -- to be related to the steadily growing supergranules, there is a clear scale separation between them and the growing supergranular flow structures. Hence, our evolutionary algorithm keeps detecting the \textit{granular} flow structures throughout the evolution of the flow. The implications on the spectrum and its time-dependent gap can be found in \ref{sec:fifteensets} and \ref{sec:different_mu}.

\begin{figure}[t!]
    \centering
    \includegraphics[scale = 1.0]{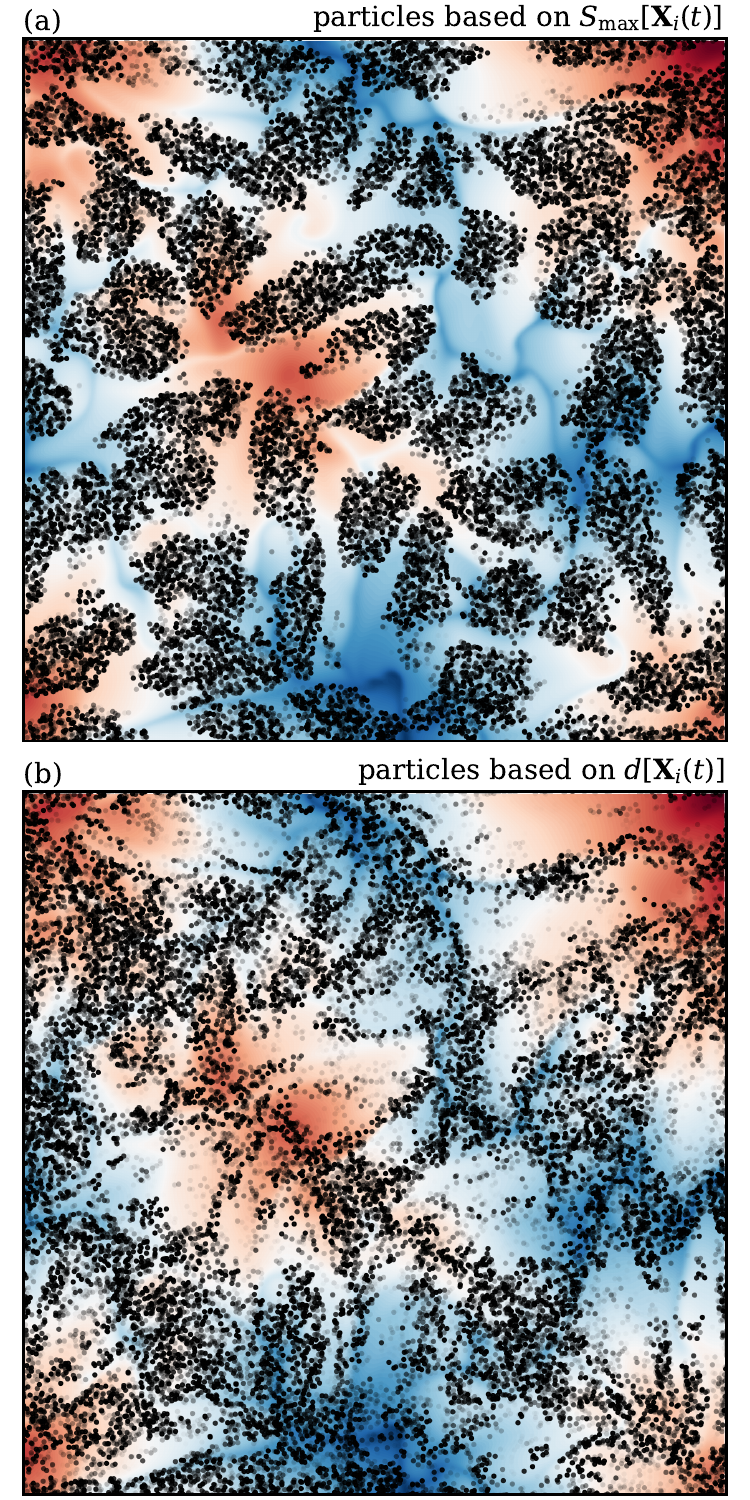}
    \caption{Relation of different measures of coherence to the growing supergranules as the thermal footprint of the second stage of the flow hierarchy. (a) Particles with $\Smax \geq S_{\textrm{max, ref}}$ do not directly correspond to the supergranule pattern (but to the granular flow structures instead). This is in contrast to particles with $d \leq d_{\textrm{ref}}$ in (b), which tend to be found more often in regions of extreme (hot or cold) temperatures. Both panels correspond to $t = 750$, see also Figure \ref{fig:smax66} (c, f, i).}
    \label{fig:superposition_T_Smax_d}
\end{figure}
Intending to further investigate the interplay between the identified coherent sets (i.e., granules) and supergranules, we contrast regions of outstanding coherence with the instantaneous temperature field in Figure \ref{fig:superposition_T_Smax_d}. In particular, we extract those trajectory fragments from Figure \ref{fig:smax66} (f) which exhibit $\Smax \left[ \bm{X}_{i} \left( t \right) \right] \geq S_{\textrm{max, ref}} \left( t \right)$ with the reference $S_{\textrm{max, ref}} \left( t \right) = \langle \Smax \left( t \right) \rangle_{N}$ -- where the right hand side denotes the time-dependent ensemble average of $\Smax \left[ \bm{X}_{i} \left( t \right) \right]$ -- and show the result in panel (a).
Although we find that the extracted coherent regions do not coincide with the large-scale supergranule pattern, finer temperature ridges -- indicating up- and down-welling fluid in the form of detaching thermal plumes -- can be found to be located between these coherent regions. This observation underlines once more the accordance of the extracted coherent sets or regions with the granular flow structures from the Eulerian frame of reference. Vice versa, one may similarly state that even the granule patterns leave a footprint in the temperature field. As such ridges are more intense when they correspond to the supergranular pattern, the evolution and location of the granular sets is consequently even influenced by the growing supergranules.

\begin{figure*}[t!]
    \centering
    \includegraphics[scale = 1.0]{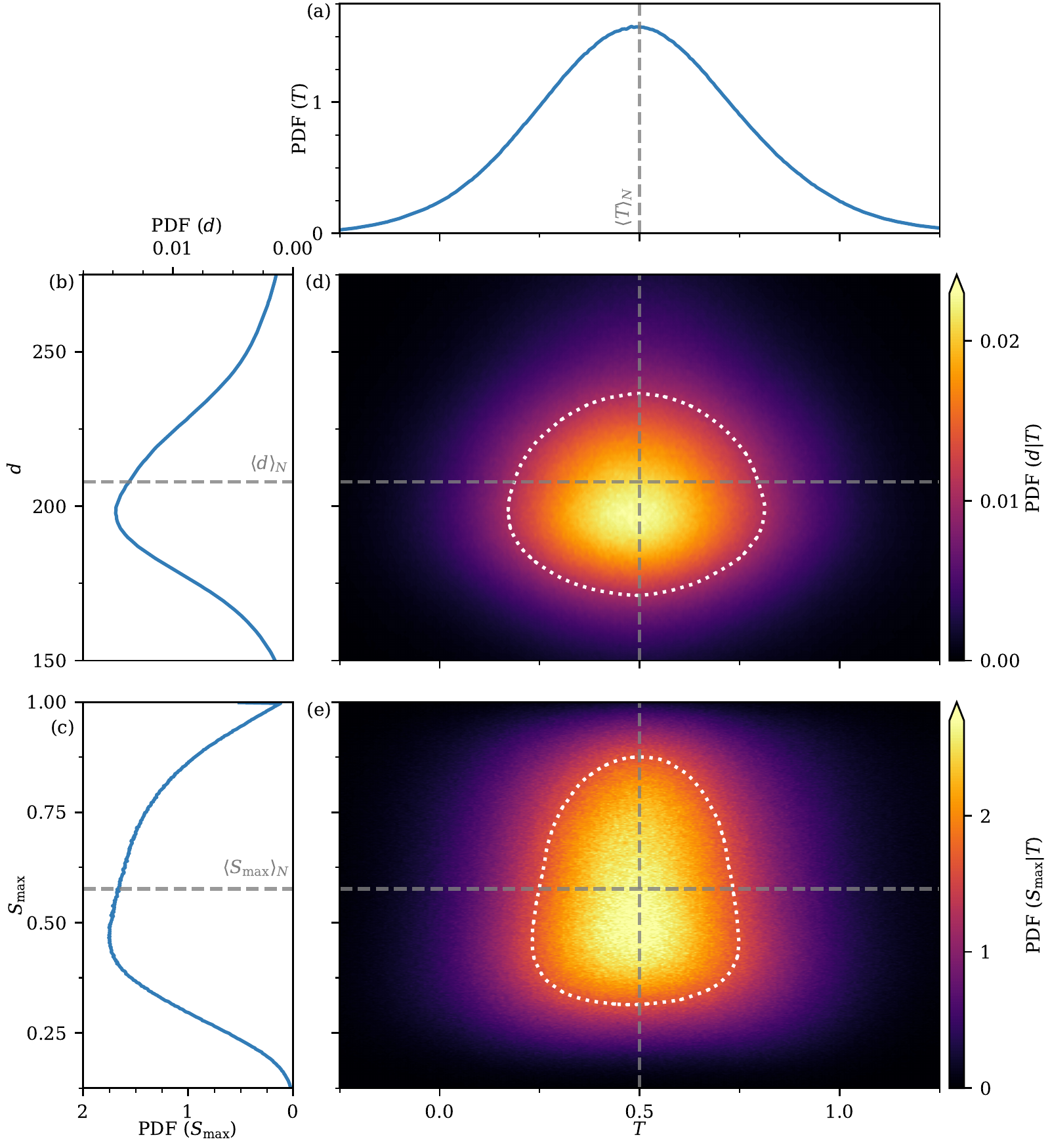}
    \caption{Statistical analysis of the temperature $T$, node degree $d$, and maximum likelihood of cluster affiliation $\Smax$. While all distributions are very wide, the extreme temperatures -- that are characteristics for the supergranules -- are found more often for below-average values of $d$ or $\Smax$ as can be deduced by the pear-shaped distributions in panels (d, e). The gray dashed lines indicate the mean values, whereas the white dotted lines enclose the joint probability density function domains surrounding the maxima that integrate up to $50 \%$. This analysis exploits the entire dataset and is thus time-independent.}
    \label{fig:probability_distributions_T_d_Smax}
\end{figure*}

This above evolutionary clustering analysis is conditioned to a certain initial number of coherent features which we selected in the first observation window based on physical arguments about the size of the granular flow structures. In other words, it is fed with additional information on the (so far granular) flow structures of interest. An alternative option to extract coherent features without such a conditioning based on user-provided information is possible via the node degree $d$.

The node degree provides local information of the structure of the Lagrangian trajectory network weight matrix and represents thus an alternative measure of coherence. Hence, we repeat our analysis of the present coherence from above but in the following based on $d$ -- the result is contrasted in Figure \ref{fig:smax66} (g -- i). This time, we find growing but less pronounced structures that seem to correlate with the thermal footprint of the supergranules -- in particular, the growing hot and cold spots, i.e., the up- and down-welling parts of the supergranules, seem to be characterized by relatively small node degrees. This first impression is further supported by Figure \ref{fig:superposition_T_Smax_d} (b) where again regions of outstanding coherence (now based on $d$) are visualised. Note here that strong coherence corresponds, in contrast to $\Smax$, to \textit{small} node degrees $d \left[ \bm{X}_{i} \left( t \right) \right] \leq d_{\textrm{ref}} \left( t \right)$ compared to the reference value $d_{\textrm{ref}} \left( t \right) = \langle d \left( t \right) \rangle_{N}$ (where the right hand side again denotes the time-dependent ensemble average of $d \left[ \bm{X}_{i} \left( t \right) \right]$).

To quantify these visual impressions, we perform a statistical analysis of the temperature $T$, node degree $d$, and maximum likelihood of cluster affiliation $\Smax$ in Figure \ref{fig:collection_PDFs}. Note that this analysis exploits the entire dataset such that the result is independent of the selected time. While panels (a -- c) show the individual distributions of the before-mentioned quantities together with their corresponding mean values, their joint probability density functions are provided in panels (d, e). Albeit all PDFs show wide distributions, the shapes of the peaks of the joint PDFs are quite similar. We highlight this pear-shaped peak via white dotted lines that surround the domain around the peak, the latter of which integrates up to $50 \%$ probability. On the one hand, as more extreme temperature values are more likely for below-average node degrees (see panel (d)), coherent trajectory fragments in the sense of $d \left[ \bm{X}_{i} \left( t \right) \right] \leq d_{\textrm{ref}}$ are confirmed to correspond more likely to the supergranular flow structures. On the other hand, coherent trajectory fragments in the sense of $\Smax \left[ \bm{X}_{i} \left( t \right) \right] \geq S_{\textrm{max, ref}}$ show more often mean temperatures (see panel (e)) and thus correspond to other flow features (i.e., the granules). Hence, both joint PDFs confirm the visual impressions from above.

These analyses based on two different measures of coherence ($\Smax$ and $d$) show that it is possible to extract both stages of the present hierarchy of long-living large-scale flow structures (granules and supergranules, respectively). Our previous studies have shown that supergranules contribute significantly to the heat transport \cite{Vieweg2021a}, whereas coherent sets at the centres of convection rolls (in the case of turbulent superstructures) reduced the overall heat transport \cite{Vieweg2021,Schneide2022}. This apparent contradiction asks to study the heat transport for the present configuration once again depending on the chosen measure of coherence.

\begin{figure*}[t!]
    \centering
    \includegraphics[scale=1.0]{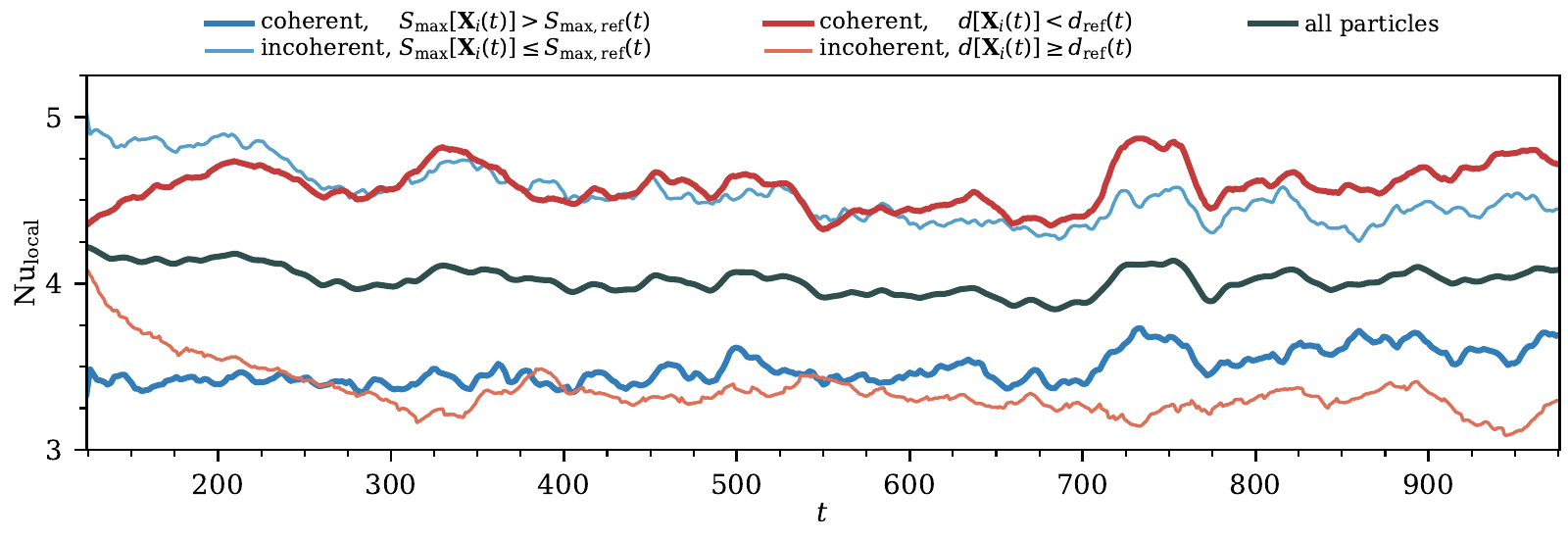}
    \caption{Contributions of coherent and incoherent regions to the global heat transport over time. Depending on the chosen measure of coherence ($\Smax$ or $d$), coherent trajectories correspond either to the granular or supergranular flow structures and exhibit thus a decreased or increased heat transport, respectively. 
    The definition of reference values for judging the coherence is described in section \ref{sec:results}.}
    \label{fig:evolution_Nu}
\end{figure*}

\begin{figure}[t!]
    \centering
    \includegraphics[scale=1.0]{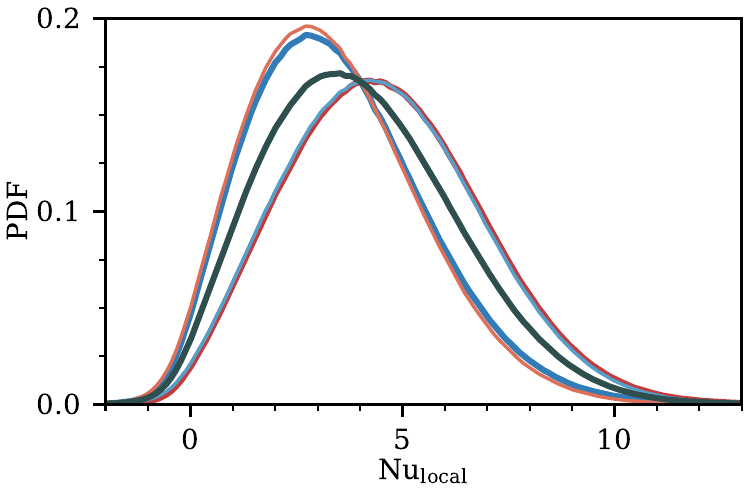}
    \caption{Probability density functions of the (partly conditioned) local Nusselt numbers $\Nuloc$ as defined in eq. \eqref{eq:localNu_av}. The color-coding coincides with the one outlined in Figure \ref{fig:evolution_Nu}. Based on $k_{\textrm{init}} = 66$ and $\mu = 0.95$, these distributions are derived from a joint analysis of all times at once. Hence, the reference values for the separation of the different groups of local Nusselt numbers are given by the (time-independent) mean values indicated in Figure \ref{fig:probability_distributions_T_d_Smax} (b, c), i.e., $S_{\textrm{max, ref}} \neq S_{\textrm{max, ref}}(t)$ and $d_{\textrm{ref}} \neq d_{\textrm{ref}}(t)$. Note that this is in contrast to Figure \ref{fig:evolution_Nu} where the separation was performed based on instantaneous reference values, i.e., $S_{\textrm{max, ref}} = S_{\textrm{max, ref}}(t)$ and $d_{\textrm{ref}} = d_{\textrm{ref}}(t)$.}
    \label{fig:collection_PDFs}
\end{figure}

\section{Conditional heat transfer analysis}
\label{sec:results1}
A statistical analysis of the heat transfer properties along trajectory segments -- coherent or incoherent, depending on the choice of the measure of coherence -- allows to scrutinize our previous results. In Figure \ref{fig:evolution_Nu}, we consider the time-dependent local Nusselt numbers averaged over different (coherent or incoherent) ensembles. The dark line represents the Nusselt number obtained from all particles (independent of their coherence) and offers $\langle \Nuloc \rangle_{N, t} = 4.02 \pm 0.07$ (indicating the mean value and standard deviation). Note that the standard deviation measures here the fluctuations in time of the time-dependent ensemble averages of the local Nusselt number which indicates that the overall Lagrangian heat transport does not change much over time. Based on the chosen measure of coherence and its time-dependent reference value ($S_{\textrm{max, ref}} \left( t \right)$ or $d_{\textrm{ref}} \left( t \right)$), the evolution of the correspondingly averaged local Nusselt numbers changes.

On the one hand, a separation based on $\Smax$ results in the two blue curves with $\langle \Nuloc \rangle_{N_{S \textrm{, coh}}, t} = 3.48 \pm 0.09$ and $\langle \Nuloc \rangle_{N_{S \textrm{, inc}}, t} = 4.53 \pm 0.16$. The local Nusselt number along trajectories in coherent regions is significantly smaller than those in mixing regions, implying that they participate only weakly in the turbulent heat transfer across the fluid layer. This is in agreement with our previous results in \cite{Schneide2022}.

On the other hand, when we consider a separation based on $d$ (yielding the red curves) we obtain $\langle \Nuloc \rangle_{N_{d \textrm{, coh}}, t} = 4.58 \pm 0.12$ and  $\langle \Nuloc \rangle_{N_{d \textrm{, inc}}, t} = 3.35 \pm 0.14$. In contrast to our previous Lagrangian studies in \cite{Schneide2022}, trajectories in coherent regions (in the sense of a node degree below average) contribute here significantly more to the vertical heat transport than those from the mixing regions. These results are confirmed when studying the time and ensemble averages which are summarized in Table \ref{tab:Nusselt_details} or the time-independent conditioned PDFs as shown in Figure \ref{fig:collection_PDFs}.

\begin{table*}[tb]
\centering
\begin{tabular}{c | c c c c}
\hline
$S_{\textrm{max, ref}} \pm \sigma_{\textrm{s}}$                 & coherent                              & incoherent                        & (strongly) coherent                                       & (strongly) incoherent \\
$0.58 \pm 0.19$                                                 & $\Smax \geq S_{\textrm{max, ref}}$    & $\Smax < S_{\textrm{max, ref}}$   & $\Smax \geq S_{\textrm{max, ref}}+\sigma_{\textrm{s}}$    & $\Smax < S_{\textrm{max, ref}}-\sigma_{\textrm{s}}$ \\
\hline
$\langle \Nuloc \rangle_{N_{\textrm{group}}, \mathbb{T}}$       & $3.49\pm 2.10$                        & $4.53 \pm 2.28$                   & $2.95 \pm 1.93$                                           & $4.98 \pm 2.30$ \\ 
$\Lambda_{\rm to,N_{\textrm{group}}}$                           & $8.7 \pm 6.6$                         & $9.9 \pm 7.1$                     & $8.2\pm 6.5$                                              &  $10.6 \pm 7.4$ \\
\hline
\hline
$d_{\textrm{ref}} \pm \sigma_{d}$                               & coherent                              & incoherent                        & (strongly) coherent                                       & (strongly) incoherent \\
$207.9\pm 30.9$                                                 & $d \leq d_{\textrm{ref}}$             & $d > d_{\textrm{ref}}$            & $d \leq d_{\textrm{ref}}-\sigma_d$                        & $d > d_{\textrm{ref}}+\sigma_d$ \\ 
\hline
$\langle \Nuloc \rangle_{N_{\textrm{group}}, \mathbb{T}}$       & $4.59\pm 2.30$        & $3.36 \pm 2.02$                   & $4.64 \pm 2.55$                                           & $2.39 \pm 1.70$\\
$\Lambda_{\rm to,N_{\textrm{group}}}$                           & $9.5 \pm 7.0$         & $9.1 \pm 6.8$                     & $9.4 \pm 7.6$                                             & $8.7\pm 6.9$\\ 
\hline
\end{tabular}
\caption{Average local Nusselt numbers $\langle \Nuloc \rangle_{N_{\textrm{group}}, \mathbb{T}}$ and turnover wavelengths $\Lambda_{\rm to,N_{\textrm{group}}}$ (both time and ensemble average) for different groups of trajectories $N_{\textrm{group}}$ -- the separation disentangles coherent and incoherent groups ($\textrm{group} = \textrm{coh}$ and $\textrm{group} = \textrm{inc}$, respectively).}
\label{tab:Nusselt_details}
\end{table*}

Convective heat transfer across the fluid layer depends crucially on turnover movements of the individual particles. Thus, we finally determine the average turnover wavelengths depending on the chosen ensemble. If this is not conditioned to certain coherent or incoherent trajectories, this yields $\Lambda_{\rm to} = 9.3 \pm 6.9$ -- note that this result differs slightly from the results outlined in section \ref{sec:problem} as we restrict our clustering analysis to only a subset of the entire dataset. Additionally, we restrict the computation to coherent and incoherent regions based on both $\Smax$ and $d$ in the same manner as for the local Nusselt number above. The results are included in Table \ref{tab:Nusselt_details} and show that the turnover wavelengths of coherent sets are smaller (than those of incoherent sets) when coherence is measured with respect to $\Smax$, whereas they are larger once coherence is measured with respect to $d$.

This quantitative study underlines the relevance of the chosen measure of coherence. On the one hand, smaller granule-sized coherent sets (identified based on $\Smax$ via spectral clustering) contribute less to the global heat transport than their complements. On the other hand, coherent trajectories characterized by a small node degree $d$ are found to belong to larger, supergranular flow structures that play a key role in the vertical heat transfer. In \cite{Vieweg2021a}, it was shown in the Eulerian frame of reference that the global Nusselt number remained practically unchanged in the course of the slow aggregation -- however, the fraction of the heat transfer due to the supergranule steadily increased until the structure filled the entire cell. The above-mentioned apparent contradiction is thus indeed consistent with our previous Lagrangian and Eulerian analyses.

\section{Conclusion and outlook}
\label{sec:conclusion}
The present work investigates a Rayleigh-Bénard convection flow that is driven by a constant heat flux at the top and bottom boundaries and thus exhibits a hierarchy of flow structures ranging from granules to supergranules. The latter structures evolve out of the small-scale turbulence in a very slow aggregation process that lasts for several thousands of free-fall times and comes to an end only when the finite domain size prohibits any further aggregation. Such hierarchies of structures are found in several much more complex flows in nature, e.g., in convection in the interior of the Sun \cite{Schumacher2020} where they are eventually constrained by rotation \cite{Schumacher2020,Rincon2018,Vieweg2022} (which is not considered here for simplicity). 

In this present study, we obtained time-dependent coherent sets within a network-based framework based on Lagrangian trajectories, i.e., massless particles which follow the surrounding time-dependent velocity field perfectly. We analysed the flow based on two different measures of coherence. 
On the one hand, we demonstrated that granular coherent structures can be identified by an evolutionary spectral clustering approach and contribute only little to the overall heat transfer. This confirms our previous results \cite{Vieweg2021,Schneide2022} that have been obtained for the complementary thermal boundary conditions, i.e., for the case of applied constant temperatures at the top and bottom planes. 
On the other hand, an analysis of the node degree provided insights on the objectively most coherent flow structures present in the flow and extracted the supergranular flow structures instead. These structures were, in accordance with our previous Eulerian studies \cite{Vieweg2021a}, found to contribute significantly to the overall heat transfer. 

This apparent contradiction is the result of the present hierarchy of different long-living large-scale flow structures -- in other words, different measures of coherence extract here different stages of the hierarchy. The spectral clustering approach based on the maximum likelihood of cluster affiliation $\Smax$ extracted the centres of convective granular flow structures (see also \cite{Vieweg2021}), whereas the technically simpler approach based on the node degree $d$ extracted the coherently up- and down-welling parts of the growing supergranular flow structures. Note that such spatial regions outside the centres of convection rolls correspond in the opposite classical case of turbulent superstructures to the incoherent background with \textit{high} node degrees \cite{Schneide2022}.
To summarize, this highlights that our evolutionary clustering algorithm is capable (i) to analyse the gradual evolution of granular Lagrangian coherent sets, and (ii) to detect the even larger transient supergranular flow structures by alternative measures of coherence. This result is promising since many convective flow phenomena are determined by such a transient evolution, see e.g. the diurnal cycle for atmospheric boundary layers.

In future work, we plan to extend the Lagrangian framework to experimental systems, especially from chemical process engineering, see e.g. \cite{Weiland2023} for a first study. In chemical reactors, the detailed knowledge of the hierarchy of coherent flow structures may help to identify zones of poor mixing and enable us to tailor the mixing and reaction processes. Mixing and entrainment processes in turbulent clouds are just another possible application of the present framework.

\appendix

\section{Comparison of different graph constructions}
\label{sec:graph_construction}

\begin{figure}[t!]
    \centering
    \includegraphics[scale=1.0]{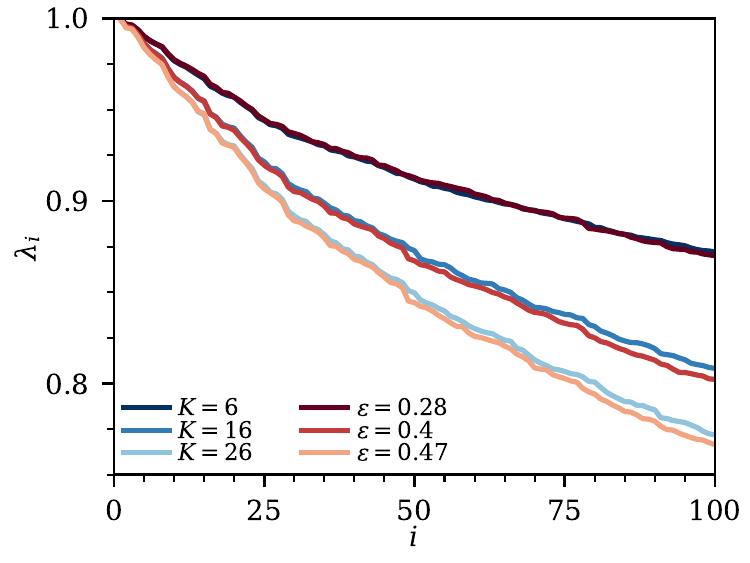}
    \caption{Initial eigenspectra obtained from different measures of graph construction. Albeit the graph can be constructed based on $K$ mutual nearest neighbours or $\varepsilon$-close encounters, the resulting spectra can be quite similar. Note that a construction based on $K = 26$ mutual nearest neighbours coincides with the approach in the main text.}
    \label{fig:comparison_spectra_K_eps}
\end{figure}
Setting up the networks by means of $K=26$ instantaneous mutual neighbours is inspired by a cubic grid consisting of $27$ grid points, with the centre point having 26 neighbours in this sense. This results in a link density of about $0.003$ for the respective network weight matrices, which is well in the range proposed by \cite{Donner2010} in the related context of recurrence networks. This choice corresponds to a mean $\varepsilon =0.47$ and compares well to our previous study in \cite{Schneide2022}. There, $\varepsilon=0.14$ was chosen for the same number of trajectories but for a smaller domain of aspect ratio $\Gamma=16$. The resulting  difference in the density of the trajectory seeding by a factor $3.5$ is directly reflected in the choice of the respective $\varepsilon$ values.
 
Figure \ref{fig:comparison_spectra_K_eps} compares the initial eigenspectra of the spectral clustering for $K=26$ and $\varepsilon=0.47$ and confirms a very good correspondence. This figure includes also the spectra of the following investigations.

In contrast, the spectra for $K=6$ (inspired by a 7-point stencil) and $\varepsilon=0.28$ (link density: $0.0007$) do not display prominent gaps in the range of the smaller eigenvalues that are of interest in our study (i.e., $50 \dots 100$ granular flow structures, see Figure \ref{fig:spectrum}) -- any clustering would result in these cases in spurious clusters (which is not shown).

The spectra for $K=16$ and $\varepsilon=0.4$ (link density: $0.0018$) are very similar to those of $K=26$ and $\varepsilon=0.47$. The corresponding clusters (not shown) compare also well, so that we can conclude that the network construction appears to be robust if sufficiently many neighbours are taken into account. Note that increasing $K$ can be interpreted as increasing diffusion, resulting in smoother clusters and a stronger decay in the spectrum.

In \cite{Filippi2021}, parameters of the spectral approach used in \cite{Hadjighasem2016} are tuned with respect to optimizing a spectral gap. This is a useful ansatz when only a smaller number of distinct coherent sets are sought. In the setting of this present paper, where we are focusing on many sets with no clear spectral gap in the region of interest, such an ansatz is not feasible.

\section{Evolutionary clustering for different $k_{\textrm{init}}$}
\label{sec:fifteensets}

\begin{figure*}[t!]
    \centering
    \includegraphics[scale=1.0]{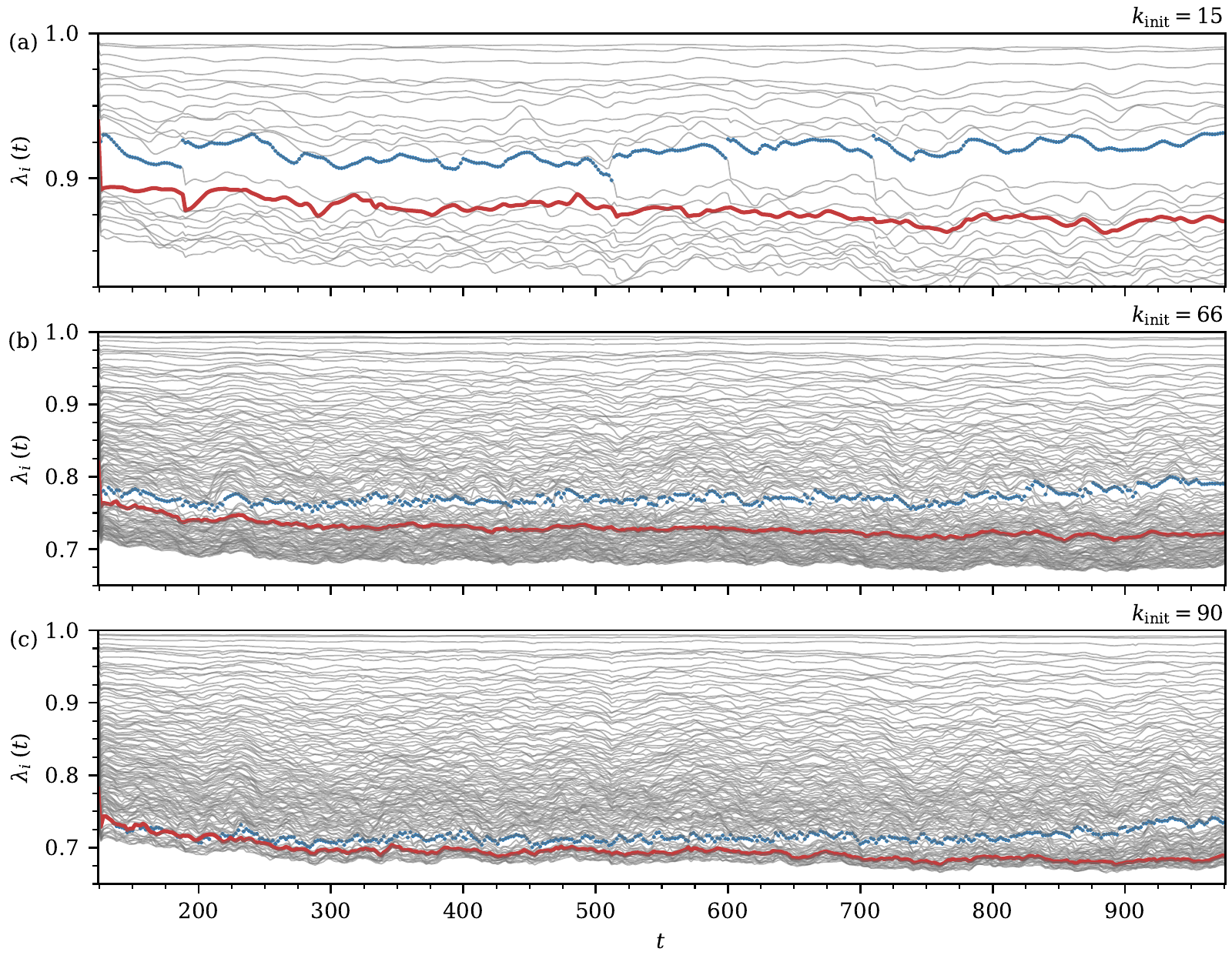}
    \caption{Evolution of eigenspectra obtained from the spectral clustering for different choices of the initial number of coherent sets $k_{\textrm{init}}$ given an obliviousness parameter of $\mu = 0.95$. The evolution of the $k_{\textrm{init}}$--th eigenvalue is highlighted in red, whereas the current gap is indicated in blue. Although the spectral clustering stabilizes the present structures, the establishing gap becomes less pronounced for higher values of $k_{\textrm{init}}$.}
    \label{fig:spectrum_evol_different_kinit}
\end{figure*}

\begin{figure*}[t!]
    \centering
    \includegraphics[scale = 1.0]{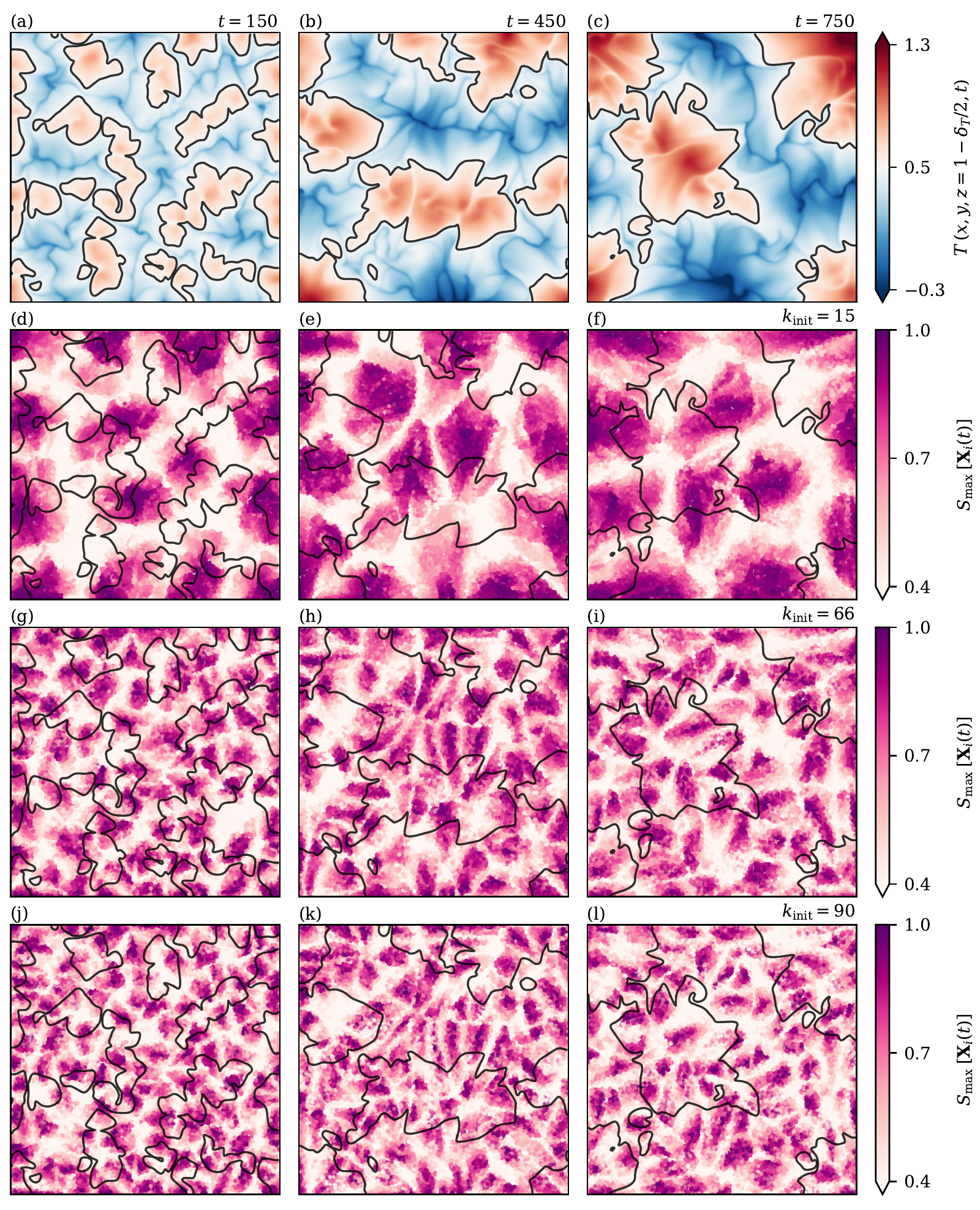}
    \caption{Comparison of the effect of the initial number of coherent sets $k_{\textrm{init}}$ on the present coherent flow features given a fixed shift of the observation window $\Delta t_{\textrm{s}} = 2$ and an obliviousness parameter $\mu = 0.95$. Note that times are chosen to coincide with Figure \ref{fig:smax66} and that the node degree from Figure \ref{fig:smax66} (g -- i) is independent of $k_{\textrm{init}}$.}
    \label{fig:evolution_Smax_mu095_different_k}
\end{figure*}

Here we present the results for different initial numbers of coherent features $k_{\textrm{init}}$ for the subsequent evolutionary clustering. In particular, we fix $\mu = 0.95$ as well as the shift of the observation window $\Delta t_{\textrm{s}} = 2$ (equal to main text) and compare $k_{\textrm{init}} \in \{15, 66, 90\}$ corresponding to the major spectral gaps seen and highlighted in Figure \ref{fig:spectrum} in the main text. The evolution of the eigenspectra is shown in Fig.\ \ref{fig:spectrum_evol_different_kinit}. Mergers of coherent sets manifest as upward-moving spectral gaps in terms of the number of eigenvalues, whereas splittings correspond to downward-moving gaps. The temporal evolution of the cluster indicator $\Smax \left[ \bm{X}_i \left( t \right) \right]$ is shown in Figure \ref{fig:evolution_Smax_mu095_different_k}.

For $k_{\textrm{init}} = 90$, the number and positions of the extracted coherent sets appear to be very similar to the case with $k_{\textrm{init}} = 66$ (as was used in the main text). There are a number of merging events visible in the evolution of the spectrum, see Figure \ref{fig:spectrum_evol_different_kinit} (b, c), but the resulting clusters again represent the granules rather than the supergranules. The statistics of heat transport related to the coherent and incoherent structures is also very similar for $k_{\textrm{init}} = 66$ and $k_{\textrm{init}} = 90$, see Table \ref{tab:Nusselt_details_K_mu}.

For $k_{\textrm{init}} = 15$, the numbers and positions of coherent sets are less dynamic. We observe spectral signatures of four merging events only, such that at the end of the monitoring $11$ structures are remaining -- the spectral gap can thus eventually be found after the $11$th eigenvalue, see Figure \ref{fig:spectrum_evol_different_kinit} (a).
The resulting incoherent regions tend to lie in the centers of the supergranules, where fluid is up- and down-welling. However, the differences in heat transport contributions by the coherent and incoherent regions are not as pronounced as for $k_{\textrm{init}} = 66$ and $k_{\textrm{init}} = 90$, see again Table \ref{tab:Nusselt_details_K_mu}. 
Consequently, even selecting $k_{\textrm{init}} = 15$ in the initial setup of the evolutionary clustering algorithm does not allow to directly study the gradual supergranule aggregation based on the maximum likelihood of cluster affiliation $\Smax$ given our choice of hyper-parameters.


\begin{table*}[tb!]
\centering
\begin{tabular}{clccc}
\hline
                        &               &                                                   & $\textrm{Nu}_{\textrm{local}}$ for    & $\textrm{Nu}_{\textrm{local}}$ for \\ 
                        &               & $S_{\textrm{max, ref}} \pm \sigma_{\textrm{s}}$   & $\Smax \left[ {\bm X}_i \left( t \right) \right] \geq S_{\textrm{max, ref}}$    & $\Smax \left[ {\bm X}_i \left( t \right) \right] < S_{\textrm{max, ref}}$ \\ 
\hline
$k_{\textrm{init}}= 15$ & $\mu = 0.8$   & $0.69 \pm 0.21$                                   & $3.71 \pm 2.19$                       & $4.45 \pm 2.28$ \\
                        & $\mu = 0.95$  & $0.62 \pm 0.22$                                   & $3.79 \pm 2.22$                       & $4.28 \pm 2.27$ \\
                        & $\mu = 1.0$   & $0.56 \pm 0.20$                                   & $3.63 \pm 2.14$                       & $4.41 \pm 2.30$ \\ 
\hline
$k_{\textrm{init}}= 66$ & $\mu = 0.8$   & $0.60 \pm 0.20$                                   & $3.51 \pm 2.11$                       & $4.54 \pm 2.28$ \\ 
                        & $\mu = 0.95$  & $0.58 \pm 0.19$                                   & $3.49 \pm 2.10$                       & $4.53 \pm 2.28$ \\
                        & $\mu = 1.0$   & $0.54 \pm 0.19$                                   & $3.35 \pm 2.05$                       & $4.63 \pm 2.27$ \\ 
\hline
$k_{\textrm{init}}= 90$ & $\mu = 0.8$   & $0.58 \pm 0.20$                                   & $3.49 \pm 2.12$                       & $4.54 \pm 2.27$ \\
                        & $\mu = 0.95$  & $0.56 \pm 0.19$                                   & $3.44 \pm 2.10$                       & $4.55 \pm 2.27$ \\ 
                        & $\mu = 1.0$   & $0.54 \pm 0.19$                                   & $3.34 \pm 2.05$                       & $4.64 \pm 2.26$ \\ 
\hline
\end{tabular}
\caption{Average local Nusselt numbers for the coherent and incoherent groups of particles measured by $\Smax$ for different choices of $k_{\textrm{init}}$ and $\mu$. We use $\left \langle\Smax\right\rangle_{N, \mathbb{T}}=S_{\textrm{max, ref}}\pm \sigma_S$ (mean and standard deviation) and identify the trajectory segments $\bm{X}_i(t)$ for which $\Smax \left[ {\bm X}_i \left( t \right) \right]$ is smaller (incoherent) or larger (coherent) than the reference value $ S_{\textrm{max, ref}}$. The analysis shows that for $k_{\textrm{init}}=15$ the difference in the local Nusselt numbers for the coherent and incoherent ensembles is less pronounced as for the larger number of clusters. All parameter choices give very similar quantitative results, so that the method appears to be quite robust.}
\label{tab:Nusselt_details_K_mu}
\end{table*}

\section{Evolutionary clustering for different $\mu$}
\label{sec:different_mu}

\begin{figure*}[t!]
    \centering
    \includegraphics[scale = 1.0]{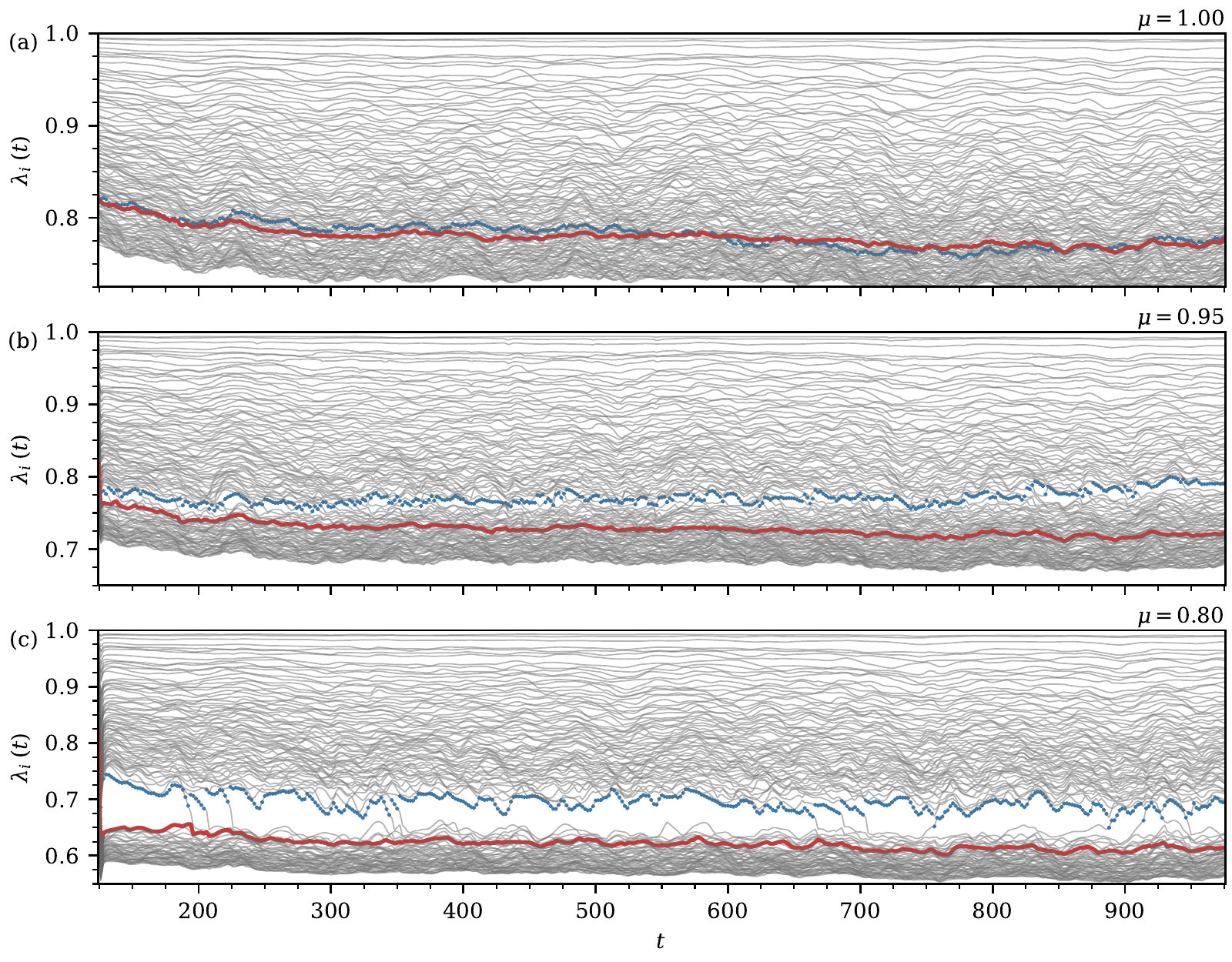}
    \caption{Evolution of eigenspectra obtained from the spectral clustering for different choices of the obliviousness parameter $\mu$ given an initial number of coherent sets of $k_{\textrm{init}} = 66$. The evolution of the $66$th eigenvalue is highlighted in red, whereas the current gap is indicated in blue. The gap becomes stronger for smaller $\mu$, stabilising the present structures even more -- thus, the gap moves less frequently.}
    \label{fig:spectrum_evol_different_mu}
\end{figure*}

\begin{figure*}[t!]
    \centering
    \includegraphics[scale = 1.0]{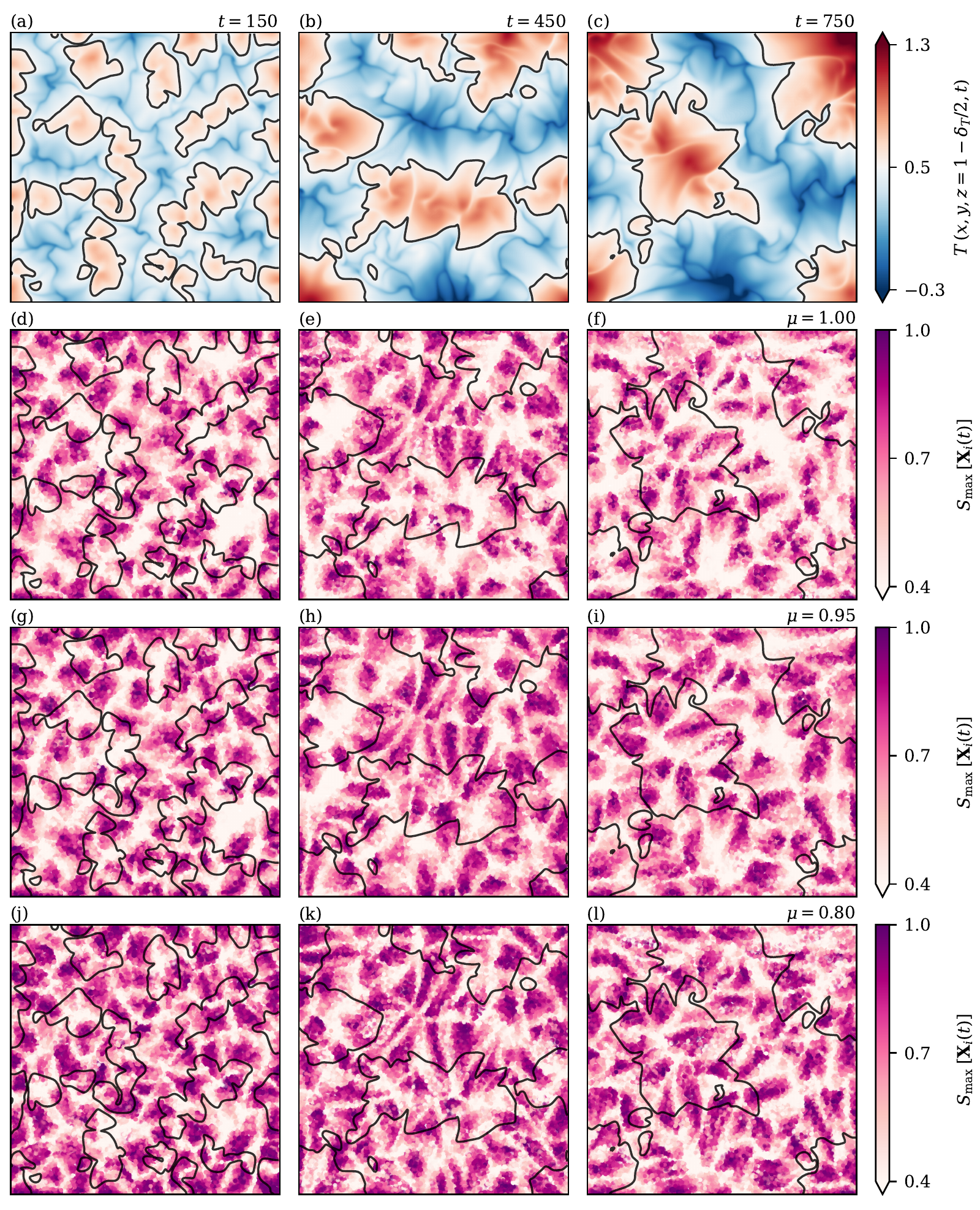}
    \caption{Comparison of the effect of the obliviousness parameter $\mu$ on the present coherent granular flow features given a fixed shift of the observation window $\Delta t_{\textrm{s}} = 2$ and $k_{\textrm{init}} = 66$ initial coherent sets. While we find that some coherent features disappear and others persist, the qualitative picture of evolving clusters is independent of the particular choice of $\mu$. Note that times are chosen to coincide with Figure \ref{fig:smax66} and that the node degree from Figure \ref{fig:smax66} (g -- i) is independent of $\mu$.}
    \label{fig:evolution_Smax_for_different_mu_k66}
\end{figure*}

Here we study the influence of the obliviousness parameter $\mu$ that weights the historical and current costs in the evolutionary clustering framework (see eq. \eqref{eq:evocl}). In particular, we fix $k_{\textrm{init}} = 66$ as well as the shift of the observation window $\Delta t_{\textrm{s}} = 2$ (equal to main text) and compare $\mu \in \{1, 0.95, 0.8\}$.
The resulting eigenspectra obtained from the spectral clustering approach are shown in Figure \ref{fig:spectrum_evol_different_mu}. As $\mu<1$ introduces a spectral gap that becomes stronger for smaller $\mu$, we observe less changes in the gaps of the spectrum when $\mu$ is decreased. This influences the stability of the clusters, as also demonstrated in Figure \ref{fig:evolution_Smax_for_different_mu_k66}. While for the choice $\mu=0.8$ more structures persist than for larger $\mu$, the qualitative picture of the evolving clusters appears to be relatively independent of the particular choice of $\mu$. This is also supported by the quantitative heat transfer analysis in Table \ref{tab:Nusselt_details_K_mu} which considers the different choices of $k_{\textrm{init}}$ and $\mu$ simultaneously.

The results of our parameter studies -- both from \ref{sec:fifteensets} and \ref{sec:different_mu} -- indicate that the coherent sets as identified in the main text are robust in the sense that neither the qualitative pictures nor the heat transfer statistics change significantly. We have fixed the length of our observation window $\Delta t_{\textrm{ow}} = 48$ on physical grounds. Small changes in $\Delta t_{\textrm{ow}}$ (not shown here) do not change the overall picture, but large changes do as the definition of coherence is crucially tied to the time scale under consideration. Our choice of $\Delta t_{\textrm{s}} = 2$ generates a considerable overlap of the subsequent time windows. We expect that decreasing the fixed shift of the observation window $\Delta t_{\textrm{s}}$ will not have a strong impact on the results due to the observed robustness. In contrast, a moderate increases in $\Delta t_{\textrm{s}}$ can be balanced by choosing a smaller $\mu$. Further detailed investigations would be possible, but are beyond the scope of this paper.


\section*{Acknowledgements}
PPV and AK are supported by the Priority Programme DFG-SPP 1881 ``Turbulent Superstructures'' of the Deutsche Forschungsgemeinschaft. The authors thank Ambrish Pandey and Christiane Schneide for their previous contributions to this subject that helped to shape the present research. The authors gratefully acknowledge the Gauss Centre for Supercomputing e.V. (https://www.gauss-centre.eu) for funding this project by providing computing time through the John von Neumann Institute for Computing (NIC) on the GCS Supercomputer JUWELS at Jülich Supercomputing Centre (JSC).

\bibliographystyle{elsarticle-num} 
\bibliography{references}

\end{document}